\documentclass[nonacm,sigconf, screen]{acmart}

\AtBeginDocument{%
  \providecommand\BibTeX{{%
    Bib\TeX}}}

\setcopyright{none} 

\renewcommand\footnotetextcopyrightpermission[1]{} 

\usepackage{mathtools}
\usepackage{graphicx}
\usepackage{textcomp}
\def\BibTeX{{\rm B\kern-.05em{\sc i\kern-.025em b}\kern-.08em
    T\kern-.1667em\lower.7ex\hbox{E}\kern-.125emX}}

\usepackage{gensymb}
\usepackage{etoolbox}
\usepackage{ifthen}
\usepackage[linesnumbered,ruled,noend]{algorithm2e}

\makeatletter%
\newboolean{maketitleAfterAllMeta}
\@ifclassloaded{acmart}%
{\setboolean{maketitleAfterAllMeta}{true}}%
{\setboolean{maketitleAfterAllMeta}{false}}%
\makeatother%

\makeatletter%
\@ifclassloaded{acmart}%
{}{}%
\@ifclassloaded{elsarticle}%
{}{}
\@ifclassloaded{llncs}%
{}{}
\@ifclassloaded{jpconf}%
{%
  \usepackage{doi}%
}{}
\@ifclassloaded{IEEEtran}%
{%
  
  \newcommand{\ifIEEEtran}[1]{#1}
}%
{%
  \newcommand{\ifIEEEtran}[1]{}
  \usepackage{flushend}
}%
\makeatother%

\SetCommentSty{algCommentSty}

\SetInd{0.5em}{0.5em}

\DontPrintSemicolon
\LinesNumbered

\makeatletter
\@ifpackageloaded{algorithm2e}{\@ifclassloaded{IEEEtran}{%
  \SetAlFnt{\small}
  \SetAlCapFnt{\small}
  \SetAlCapNameFnt{\small}
}}%
\makeatother

\usepackage[utf8]{inputenc}

\usepackage[T1]{fontenc}

\usepackage{microtype}

\usepackage{relsize}
\usepackage[10pt]{moresize}

\makeatletter%
\@ifclassloaded{acmart}{}{\usepackage{amssymb}}
\makeatother%

\usepackage{siunitx}
\usepackage{nicefrac}
\usepackage{xfrac}

\usepackage{xspace}

\usepackage[normalem]{ulem}

\makeatletter%
\@ifclassloaded{acmart}{}%
{\usepackage[usenames,dvipsnames,table]{xcolor}}
\makeatother%

\makeatletter%
\@ifclassloaded{acmart}{}
{\@ifclassloaded{elsarticle}{}
{\usepackage{cite}}}
\makeatother%

\usepackage{setspace}

\usepackage{verbatim}

\usepackage{cases}

\usepackage{tabto}

\usepackage{booktabs}
\usepackage{makecell}
\usepackage{multicol}
\usepackage{multirow}
\usepackage{amsmath}
\usepackage{amssymb}
\usepackage{tikz}

\usepackage{wrapfig}
\usepackage{stfloats}
\usepackage[export]{adjustbox}

\makeatletter
\@ifclassloaded{IEEEtran}{%
  \let\MYcaption\@makecaption%
  \usepackage[font=footnotesize]{subcaption}%
  \let\@makecaption\MYcaption%
}{%
  \usepackage{subcaption}%
}%
\makeatother

\usepackage[textsize=tiny]{todonotes}

\makeatletter%
\@ifclassloaded{IEEEtran}{}{}%
\usepackage[inline]{enumitem}
\makeatother%

\usepackage{quoting}

\makeatletter%
\@ifclassloaded{acmart}{}%
{%
}%
\makeatother%

\usepackage{listings}
\usepackage{color}
\usepackage{mhchem}

\makeatletter%
\@ifpackageloaded{hyperref}%
{}%
{%
  \usepackage[hyphens]{url}
  \usepackage[breaklinks=true,bookmarks=false]{hyperref}
  \hypersetup{
    colorlinks=false,%
    pdfborder = 0 0 0
  }
}%
\makeatother%

\usepackage[capitalize,noabbrev]{cleveref} 
\newcommand*\circleb[1]{\tikz[baseline=(char.base)]{
            \node[shape=circle,fill,inner sep=0.5pt] (char) {\textcolor{white}{#1}};}}

\usepackage{stackengine}
\usepackage{graphicx}
\usepackage[frozencache,cachedir=.]{minted}
 \definecolor{bg}{rgb}{0.97,0.97,0.97}
\setminted{fontsize=\footnotesize,linenos,baselinestretch=.97,xleftmargin=6pt,numbersep=3pt,mathescape=true,escapeinside=||,bgcolor=bg}
\usepackage{eqparbox}

\usepackage{tcolorbox}
\newcounter{rulecounter}
\newcommand*{\rulegen}[1]{%
  \stepcounter{rulecounter}%
            {\textbf{Observation~\arabic{rulecounter}}}%
}

\DeclareMathAlphabet{\mathpzc}{OT1}{pzc}{m}{it}

\makeatletter
\newcommand*{\textoverline}[1]{$\overline{\hbox{#1}}\m@th$}
\makeatother

\newcommand{\NOTE}[1]{\phantom{}\begingroup\relax\ifmmode\boldmath\else\bfseries\fi\color{Cerulean}\ignorespaces#1\ignorespaces\endgroup}
\newcommand{\TODO}[1]{\phantom{}\begingroup\relax\ifmmode\else\sffamily\fi\color{BurntOrange}\ignorespaces#1\ignorespaces\endgroup}
\newcommand{\FIXME}[1]{\phantom{}\begingroup\relax\ifmmode\boldmath\else\bfseries\sffamily\fi\color{Red}\ignorespaces#1\ignorespaces\endgroup}
\newcommand{\FIXED}[1]{\phantom{}\begingroup\relax\ifmmode\else\sffamily\fi\color{Green}\ignorespaces#1\ignorespaces\endgroup}
\newcommand{\DELETE}[1]{\phantom{}\begingroup\relax\ifmmode\else\sffamily\fi\color{Red}\ifmmode\text{\sout{\ensuremath{#1}}}\else\sout{\ignorespaces#1\ignorespaces}\fi\endgroup}

\providecommand{\linesref}[2]{\hyperref[#1]{Lines~\ref*{#1}--\ref*{#2}}}
\providecommand{\Linesref}[2]{\hyperref[#1]{Lines~\ref*{#1}} \hyperref[#2]{and~\ref*{#2}}}
\providecommand{\lineref}[1]{\hyperref[#1]{Line~\ref*{#1}}}
\providecommand{\Lineref}[1]{\hyperref[#1]{Line~\ref*{#1}}}

\newcommand{\codecircle}[1]{%
    \tikz[baseline=(char.base), remember picture, overlay]{%
        \node[circle, fill=black, inner sep=2pt, outer sep=0pt, text=white, font=\footnotesize] (char) {#1};%
    }%
}

\begin{document}

\title[]{Optimizing Data Distribution and Kernel Performance for Efficient Training of Chemistry Foundation Models: A Case Study with MACE 
}

\newcommand{\myAuthorFn}{\fontsize{9.92}{12}\selectfont}
\author{Jesun Firoz}
\affiliation{
  \institution{\myAuthorFn Pacific Northwest National Laboratory}
  \country{\myAuthorFn USA}
}
\email{jesun.firoz@pnnl.gov}

\author{Franco Pellegrini}
\affiliation{
  \institution{\myAuthorFn SISSA}
  \country{\myAuthorFn Italy}
}
\email{pellefra@sissa.it}

\author{Mario Geiger}
\affiliation{
  \institution{\myAuthorFn NVIDIA}
  \country{\myAuthorFn USA}
}
\email{mgeiger@nvidia.com}

\author{Darren Hsu}
\affiliation{
  \institution{\myAuthorFn NVIDIA}
  \country{\myAuthorFn USA}
}
\email{dahsu@nvidia.com}

\author{Jenna A. Bilbrey}
\affiliation{
  \institution{\myAuthorFn Pacific Northwest National Laboratory}
  \country{\myAuthorFn USA}
}
\email{Jenna.Pope@pnnl.gov}

\author{Han-Yi Chou}
\affiliation{
  \institution{\myAuthorFn NVIDIA}
  \country{\myAuthorFn USA}
}
\email{hanyic@nvidia.com}

\author{Maximilian Stadler}
\affiliation{
  \institution{\myAuthorFn NVIDIA}
  \country{\myAuthorFn USA}
}
\email{mstadler@nvidia.com}

\author{Markus Hoehnerbach}
\affiliation{
  \institution{\myAuthorFn NVIDIA}
  \country{\myAuthorFn USA}
}
\email{mhohnerbach@nvidia.com}

\author{Tingyu Wang}
\affiliation{
  \institution{\myAuthorFn NVIDIA}
  \country{\myAuthorFn USA}
}
\email{tingyuw@nvidia.com}

\author{Dejun Lin}
\affiliation{
  \institution{\myAuthorFn NVIDIA}
  \country{\myAuthorFn USA}
}
\email{dejunl@nvidia.com}

\author{Emine Kucukbenli}
\affiliation{
  \institution{\myAuthorFn NVIDIA}
  \country{\myAuthorFn USA}
}
\email{ekucukbenli@nvidia.com}

\author{Henry W. Sprueill}
\affiliation{
  \institution{\myAuthorFn Pacific Northwest National Laboratory}
  \country{\myAuthorFn USA}
}
\email{henry.sprueill@pnnl.gov}

\author{Ilyes Batatia}
\affiliation{
  \institution{\myAuthorFn University of Cambridge}
  \country{\myAuthorFn UK}
}
\email{ib467@cam.ac.uk}

\author{Sotiris S. Xantheas}
\affiliation{
  \institution{\myAuthorFn Pacific Northwest National Laboratory}
  \country{\myAuthorFn USA}
}
\email{Sotiris.Xantheas@pnnl.gov}

\author{MalSoon Lee}
\affiliation{
  \institution{\myAuthorFn Pacific Northwest National Laboratory}
  \country{\myAuthorFn USA}
}
\email{MalSoon.Lee@pnnl.gov}

\author{Chris Mundy}
\affiliation{
  \institution{\myAuthorFn Pacific Northwest National Laboratory}
  \country{\myAuthorFn USA}
}
\email{chris.mundy@pnnl.gov}

\author{Gabor Csanyi}
\affiliation{
  \institution{\myAuthorFn University of Cambridge}
  \country{\myAuthorFn UK}
}
\email{gc121@cam.ac.uk}

\author{Justin S. Smith}
\affiliation{
  \institution{\myAuthorFn NVIDIA}
  \country{\myAuthorFn USA}
}
\email{jusmith@nvidia.com}

\author{Ponnuswamy Sadayappan}
\affiliation{
  \institution{\myAuthorFn University of Utah}
  \country{\myAuthorFn USA}
}
\email{u6024862@gcloud.utah.edu}

\author{Sutanay Choudhury}
\affiliation{
  \institution{\myAuthorFn Pacific Northwest National Laboratory}
  \country{\myAuthorFn USA}
}
\email{Sutanay.Choudhury@pnnl.gov}


\begin{abstract}

Chemistry Foundation Models (CFMs) that leverage Graph Neural Networks (GNNs) operating on 3D molecular graph structures are becoming indispensable tools for computational chemists and materials scientists. These models facilitate the understanding of matter and the discovery of new molecules and materials. 
In contrast to GNNs operating on a large homogeneous graphs, GNNs used by CFMs  
process a large number of geometric graphs of varying sizes, requiring different optimization strategies than those developed for large homogeneous GNNs.
This paper presents 
optimizations for two critical phases of CFM training: data distribution and model training, targeting MACE - a state-of-the-art CFM. We address the challenge of load balancing in data distribution 
by formulating it as a multi-objective bin packing problem. We propose an iterative algorithm that provides a highly effective, fast, and practical solution, ensuring efficient data distribution. For the training phase, we identify symmetric tensor contraction as the key computational kernel in MACE and optimize this kernel to improve the overall performance. Our combined approach of balanced data distribution and kernel optimization significantly enhances the training process of MACE. Experimental results demonstrate a substantial speedup, reducing per-epoch execution time for training from 12 to 2 minutes on 740 GPUs with a 2.6M sample dataset. 
\end{abstract}

\maketitle
\renewcommand{\shortauthors}{Jesun Firoz et al.}


\section{Introduction}


Atomistic simulation serves as the cornerstone of modern computational chemistry and materials science.
For these applications, highly accurate quantitative calculations of molecular and material properties are essential.  Density Functional Theory (DFT), a quantum mechanical ab-initio (or first-principles) approach, has long been the preferred method for such calculations due to its relatively high generality and accuracy. However, the high time complexity and significant computational resources needed for DFT calculations limit simulations to nanosecond timescales and nanometer length scales ~\cite{kulichenko2021rise, pan2021scaling}. Recently, machine learning interatomic potentials (MLIPs), computational models to map atomic environments to energies and forces, have emerged as attractive alternatives to DFT calculations~\cite{kulichenko2021rise,ko2023recent, wang2024machine}. The time-to-solution of these surrogate models is significantly shorter while their accuracy is comparable to that of DFT \cite{kaser2023neural}.

\begin{figure}[htb]
  \centering
  \includegraphics[width=\linewidth]{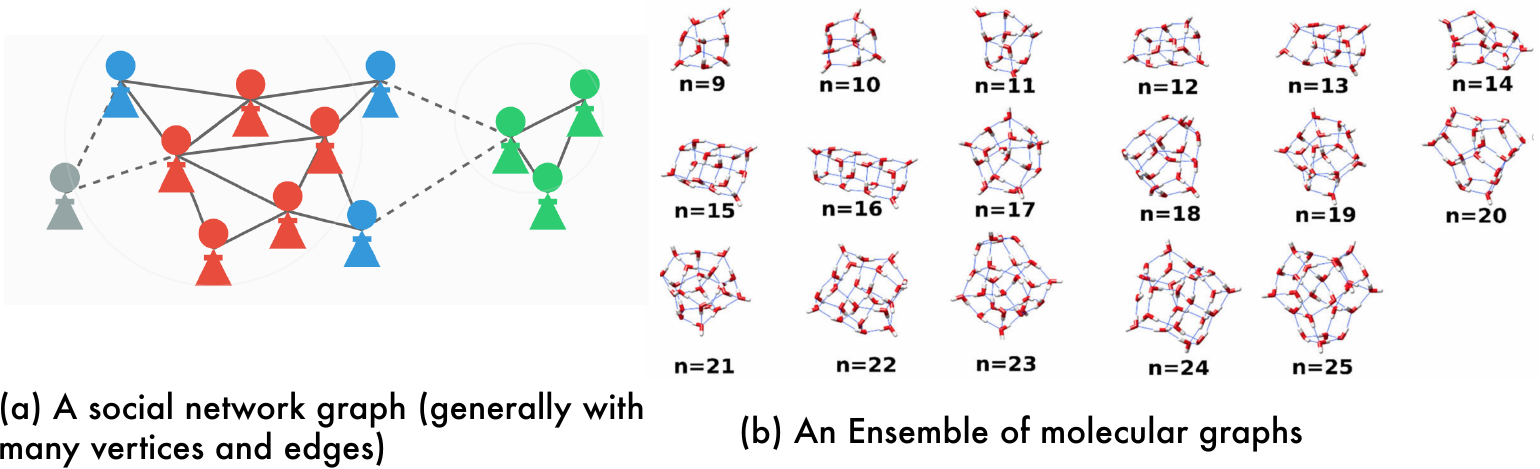}
  \caption{\small Key contrast between different classes of GNNs: Molecular GNNs are trained on many small graphs instead of a single big graph.
  }  

  \label{fig:big_vs_molecular_gnn}
\end{figure}

While traditional specialized surrogate models excel at predicting properties for \emph{specific} chemical systems, the recently-emerging \textit{foundation models} are demonstrating broad capabilities in predicting atomistic properties such as potential energy, forces, stress, and charges for a \emph{diverse} set of chemical species~\cite{bilbrey2020look, batatia2023foundation,allen2023learning,shoghi2023molecules,pasini2024scalable,zubatyuk2021teaching}. Foundation models, \textit{trained on large and diverse datasets simultaneously}, are capable of performing \textit{a wide range of general tasks}~\cite{bommasani2021opportunities}. 
The objective of a chemistry foundation model (CFM) is to unify training over a large collection of molecular and materials datasets to improve zero-shot accuracy and facilitate fine-tuning on cases with limited data. When used in zero-shot setting, CFMs can handle a wider range of chemical species than specialized models. During fine-tuning, CFMs reduce the need for extensive data collection and retraining for each new task by leveraging previously learned knowledge from diverse chemical domains \cite{niblett2024transferability}. %

Graph Neural Networks (GNNs) are the dominant architecture for chemistry foundation models (CFMs), with equivariant GNNs favored for their ability to preserve rotational symmetries in atomic systems \cite{satorras2021n, batzner20223}. However, efficiently capturing many-body interactions (meaning how
many neighboring atoms/vertices are considered simultaneously to
compute contribution to per atom energy) —a critical aspect of modeling chemical systems—remains a challenge due to the high computational costs of operations on higher-order tensors.  To address this, the atomic cluster expansion (ACE) formalism~\cite{drautz2019atomic} was developed to describe many-body interactions efficiently. Building on ACE, Batatia et al.~\cite{batatia2022mace} introduced MACE, which integrates these interactions into equivariant message-passing schemes, improving computational efficiency without sacrificing accuracy. MACE extends beyond pairwise interactions to capture complex multi-atom effects and has been widely adopted as the state-of-the-art CFM for molecular and materials modeling \cite{batatia2022mace, batatia2023foundation, kovacs2023evaluation, batatia2025design}.

\textbf{Key Traits of CFM Training Workloads.}  CFMs present distinctly different scalability challenges compared to conventional GNN workloads (cf. \Cref{fig:big_vs_molecular_gnn} and \Cref{tab:gnn_comparison}), which typically focus on single massive graph~\cite{gong2024gnnone,li2024coordinated,xia2024scaling}. Instead, CFMs train on collections of small, irregular 3D molecular graphs, requiring careful data distribution to avoid load imbalance and straggler effects across GPUs. 
These molecular graphs vary in size, sparsity, and chemical composition, making efficient batching and workload partitioning a nontrivial problem.

Another fundamental trait of CFM workloads is the dominance of symmetric tensor contractions—high-cost operations central to equivariant GNNs. These tensor operations scale poorly without specialized optimizations, making them a primary bottleneck in large-scale training. Given that CFM training can require anywhere from $\approx$3000~\cite{batatia2023foundation} to $\approx$34000 GPU-hours~\cite{shoghi2023molecules}, optimizing tensor computations and load balancing is critical for scalability.

\textbf{Why is MACE a representative CFM workload?} Many state-of-the-art equivariant GNNs, including MACE and NequIP~\cite{batzner20223}, share a common mathematical foundation, where different models correspond to specific parameter choices within a unified design space~\cite{batatia2025design}. This commonality highlights shared computational challenges, particularly the high cost of symmetric tensor contractions, making MACE an ideal target for optimization.

Beyond its architectural significance, MACE has demonstrated remarkable generalization across diverse chemical systems. Unlike traditional machine-learned force fields that require system-specific training, MACE has been successfully applied to a broad range of atomistic simulations, including solids, liquids, gases, chemical reactions, and interfaces~\cite{batatia2023foundation,batatia2022mace,batatia2025design}. This broad applicability positions MACE as a strong candidate for a foundation model in chemistry, reinforcing the importance of optimizing its computational efficiency. Therefore, by focusing on MACE, we ensure that our optimizations extend to a wider class of CFMs built on equivariant GNNs, making them broadly impactful for large-scale atomistic modeling.

{
\begin{table*}
\footnotesize
\begin{tabular}{p{0.15\textwidth}p{0.35\textwidth}p{0.35\textwidth}}
\toprule
\textbf{Aspect} & \textbf{GNNs for large Graphs} & \textbf{GNNs for Geometric/Molecular Graphs} \\
\midrule
Input graph structure & Single massive graph 
& Collection of many smaller graphs 
\\
\hline
Node count &	Millions to billions &	Typically <1000 per graph \\
\hline
Data distribution & Usually a single connected graph that needs to be partitioned across workers (partition vertices and edges) & Natural partitioning as each molecular graph is independent. Each worker processes multiple graphs in parallel. \\
\hline
Node features & Typically scalar features only & Mix of scalar (atom type) and geometric features  \\
\hline
Symmetries & Only permutation symmetry needs to be preserved & Multiple symmetries: permutation, rotation, translation \\
\hline
Edge definition & Fixed edges based on relationships & Dynamic edges often based on distance cutoffs between atoms \\
\hline
Key computations & Graph partitioning, neighborhood sampling, feature caching
& Tensor product and contraction, message passing
\\
\hline

Performance bottleneck & Communication overhead of node features between workers  & Tensor computations and geometric feature calculations \\
\bottomrule
\end{tabular}
\caption{\small Comparison between GNNs for large graphs vs. geometric/molecular graphs.}
\label{tab:gnn_comparison}
\vspace*{-3.2em}
\end{table*}
}

\textbf{Contributions.} In summary, we make the following contributions in this paper.
\begin{itemize}[noitemsep,topsep=0pt,leftmargin=*]
    \item We formulate the problem of efficient data distribution of 3D molecular graphs as a multi-objective bin packing problem with two objectives: achieving balanced data distribution while reducing the memory footprint of padding. We propose a practical, fast, and effective iterative algorithm to tackle this problem (\Cref{sec:bin_packing_load_balancer}).
    \item We identify the key computational kernel, symmetric tensor contraction,  
    that are common operations across multiple equivariant GNNs, and propose optimization techniques for the symmetric tensor contraction kernel (\Cref{sec:sym_cont_channeL_prod}).

    \item We train a GNN-based CFM on a diverse dataset of 2.6M molecular configurations, spanning 1–768 atoms and varying sparsity profiles, to study the impact of non-uniform dataset composition on load balancing and training efficiency (\Cref{sec:dataset}).
    \item Our evaluation demonstrates a speedup of 6$\times$ for the state-of-the-art CFM MACE on 740 GPUs with the dataset  containing 2.6M molecular structures, with wide range of diverse molecules, both in terms of the size and sparsity pattern (\Cref{sec:eval}).
\end{itemize}

\section{Background}\label{sec:background}

Molecules can be naturally represented as 3D geometric graphs, where atoms are vertices and bonds are edges, with additional geometric attributes such as atomic coordinates, angles etc. Therefore, molecular GNNs need to account for spatial information such as atomic coordinates and distances, and respect physical symmetries. Molecular structures can be altered through rotations, reflections, or translations without changing their fundamental structure or properties. \textit{Invariant GNNs}~\cite{schutt2017schnet,gasteigerdirectional,gasteiger2021gemnet,coors2018spherenet} produce outputs that are independent of the molecule's orientation, while \textit{equivariant GNNs} (EGNNs)~\cite{batzner20223,batatia2022mace,thomas2018tensor,musaelian2023learning,liaoequiformer} preserve the relationship between input features and their corresponding hidden representations under symmetry transformations like rotations.  
Also, for molecular GNNs the message passing mechanism is adapted to incorporate both topological and geometric attributes.  The next two sub-sections provide necessary background on these two key aspects.

\subsection{Equivariant Molecular Representations: Motivation}

In molecular modeling, atoms interact in three-dimensional space, and their properties should change predictably when the molecule is rotated. This is crucial for tasks like energy prediction and force calculations, where rotational symmetry plays a fundamental role.  Equivariant representations ensure that a model’s understanding of atomic environments transforms correctly under rotation—preserving the underlying physics. Models such as MACE learns equivariant representations using the mathematical framework of group theory and spherical harmonics to describe atomic environments while naturally handling rotations. 

\textit{Group theory} ~\cite{zee2016group} provides a mathematical framework for describing \textit{symmetries} in physical systems. In the context of molecular modeling, the \texttt{SO(3)} group represents all possible 3D rotations. A \textit{representation} of \texttt{SO(3)} describes how objects transform under these rotations, and \textit{irreducible representations} (\texttt{irreps}) are the fundamental building blocks that cannot be decomposed further.

\textit{Spherical harmonics} are special mathematical functions that describe how properties vary on the surface of a sphere and  describe how atoms interact in 3D space while naturally preserving rotational properties. Spherical harmonics are typically denoted as $Y_l^m(\mathbf{\hat{r}})$, where $l$ is the \textit{degree/order} $(0, 1, 2, \cdots)$, $m$ is the \textit{azimuthal order/phase} $(-l \leq m \leq l)$
and $\mathbf{\hat{r}}$ is a vector lying on the unit sphere. Spherical harmonics of degree $l$ form a basis for the $(2l+1)$-dimensional irreducible representation of \texttt{SO(3)}, the 3D rotation group. This property makes them ideal for constructing rotationally equivariant neural network layers. 


\begin{figure*}[htbp]
  \centering
  \vspace*{-1.2em}
  \includegraphics[width=0.8\textwidth]{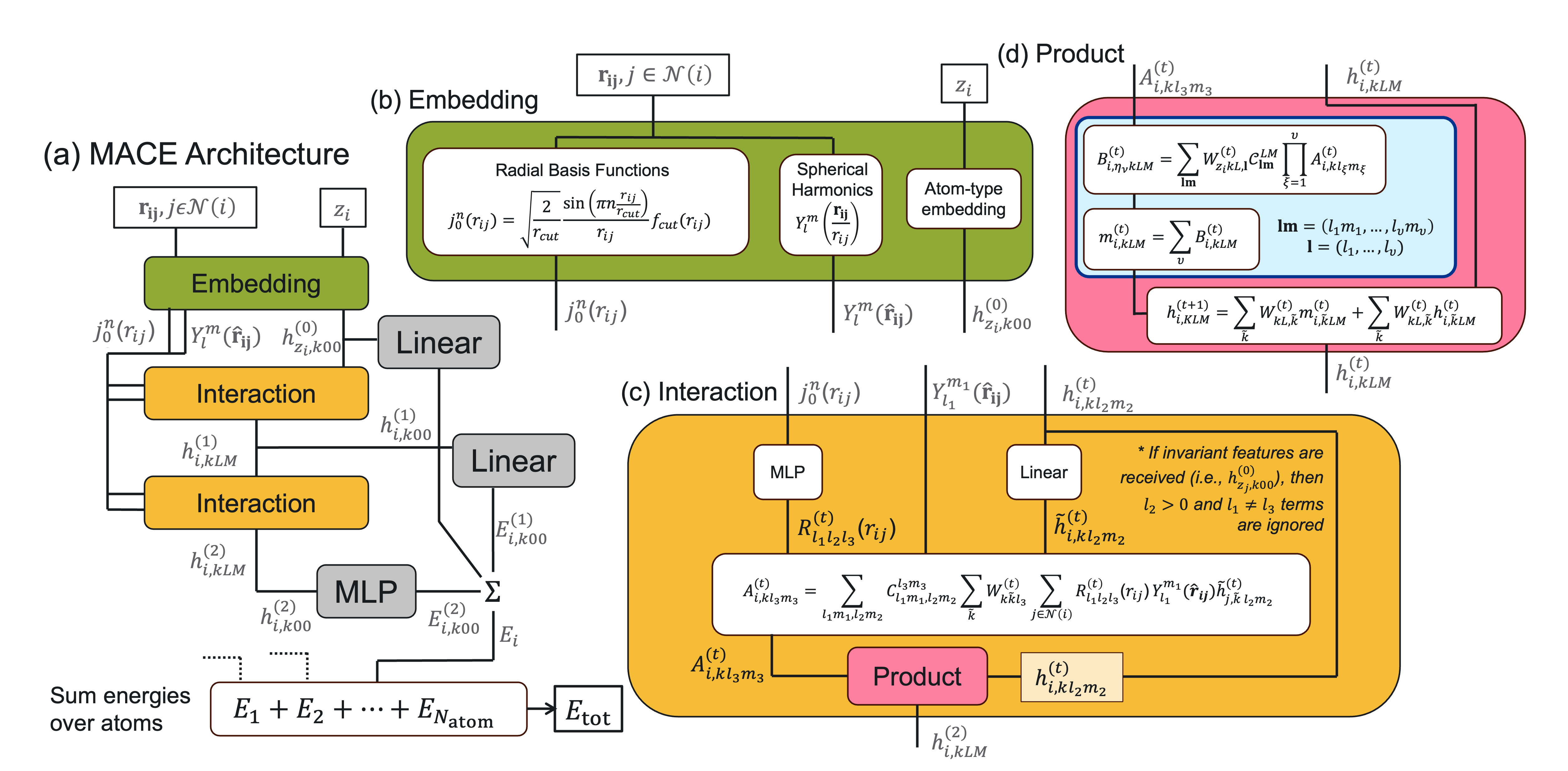}
\vspace*{-1.2em}
\caption{\small A schematic diagram of the MACE model (a-d) \cite{batatia2022mace}. In (a), input and output tensor are shown for each of the embedding, interaction, linear readout, and MLP readout layers. Two interaction layers are shown as this is the number typically used in MACE. The construction of the $A_{i,kl_3m_3}$ is shown in the interaction layer (c) with details of the higher body-order products shown in (d). \Cref{alg:mace_contraction} shows the details for computing the product in the blue box of (d).}
\label{fig:model_architecture_symmtric_contraction}
\vspace*{-1.5em}
\end{figure*}

\vspace{-1em}
\subsection{Learning Equivariant Representations via Message-Passing Architecture}

MACE~\cite{batatia2022mace} is an equivariant message-passing GNN where each layer encodes many-body information of atomic geometry. A schematic diagram of the MACE model is shown in \Cref{fig:model_architecture_symmtric_contraction} (a-d). In MACE, each atom is associated with its position $\mathbf{r_i}$, the atomic numbers $z_i$, and a set of equivariant learnable features $h_{i,klm}$, where $k$ denotes the learnable channel, $l$ and $m$ relates to the order of the spherical harmonics.

First, the \textit{embedding layer} (\Cref{fig:model_architecture_symmtric_contraction}(b)) creates a set of features for the nodes and edges describing the chemical species and the length and orientation of each edge. This is done by encoding the interatomic displacements ($\mathbf{r_{ji}}$) in a set of radial basis functions and spherical harmonics $Y_l^m(\mathbf{\hat{r}_{ji}})$, where $\mathbf{\hat{r}_{ji}}$ is the unit vector that points from a neighboring atom $j$ to atom $i$. 

Next, the \textit{interaction layer} pools information from the neighbors (\Cref{fig:model_architecture_symmtric_contraction}(c)) and computes per-atom features $A_{i,klm}$ of the surrounding environment from the radial $R_{l_1, l_2, l_3}(r_{ji})$, spherical $Y_l^m(\mathbf{\hat{r}_{ji}})$, and per atom $h_{iklm}$ embeddings, while maintaining equivariance of these features using the Clebsch-Gordan Coefficients.

\begin{table}[t]
\caption{List of symbols and notation}\label{tab:notations}
\vspace{-1em}
\footnotesize
\begin{tabular}{@{}ll@{}}
\toprule
Symbol & Description \\
\midrule
$Y^m_{ji,l}$ & Spherical harmonics of order $l$, component $m$ \\
& for edge vector $ji$ \\
$\mathbf{r}_{ji}$ & Interatomic displacement vector between atoms \\
& $j$ and $i$ \\
$\hat{\mathbf{r}}_{ji}$ & Unit vector pointing from atom $j$ to atom $i$ \\
$h^{(t)}_{j,klm}$ & Learnable equivariant features at iteration $t$ for atom $j$, \\
& channel $k$,  order $l$, component $m$ \\
$A_{i,klm}$ & Atomic basis features \\
$C^{l_3m_3}_{l_1m_1,l_2m_2}$ & Clebsch-Gordan coefficients for combining spherical \\ & harmonics of orders $(l_1,m_1)$ and 
 $(l_2,m_2)$ into \\ & order $(l_3,m_3)$ \\
$C^{LM}_{lm}$ & Generalized Clebsch-Gordon coefficients \\
$R^{(t)}_{ji,kl_1l_2l_3}$ & Weights at iteration $t$ for edge $ji$, channel $k$, and \\ 
& spherical harmonic orders $l_1$, $l_2$, $l_3$ \\
$W^{(t)}_{z_ikL,l}$ & Learnable weights at iteration $t$ dependent on atomic \\ & species $z_i$, channel $k$, orders $L,l$ \\
$z_i$ & Atomic numbers \\
$\nu_{\text{max}}$ & Maximum correlation order \\
$L_{\text{max}}$ & Maximum irreducible representation order \\
\bottomrule
\end{tabular}
\end{table}

The \textit{product layer} (\Cref{fig:model_architecture_symmtric_contraction} (d)) then  raises the information aggregated from the neighbors to a power to create high body-order features $B_{i,\eta_\nu,kLM}$ from the $A_{i,klm}$ features. This is done by taking a product of the $A_{i,klm}$ up to a correlation order $\nu$. This product mixes the various equivariance levels of $A_{i,klm}$, each encoding different aspects of the local chemical environment, to produce a feature that encodes up to $\nu$ atom interactions. 

Following the framework of message passing GNN, message at node $i$ for  iteration step $t$ is then constructed as 
\begin{align}
m_i^{(t)} = \bigoplus_{j \in N(i)} M_i(\sigma_i^{(t)}, \sigma_j^{(t)})
\end{align}
where $N(i)$ is the neighborhood of node $i$, $M_i$ is a learnable message function, and $\bigoplus_{j \in N(i)}$ is the \textit{aggregation} operation, $\sigma_i$ is the state of node $i$. In MACE, for channel $k$, spherical harmonic order $L, M$ and learnable weights $W$, the message is computed as
\begin{align}
m_{i,k,L,M}^{(t)} = \sum_\nu \sum_{\eta_i} W_{k,\eta_i}^{(\nu)} B_{\eta_i,k,L,M}^{(\nu,s)}    
\end{align}
The inner sum considers all possible combinations of $(l,m)$ and this is done up to a certain body order $\nu$. The per-atom features are then updated using a residual update from these higher body-order features and the previous $h_{i,klm}$. 

At each intermediate interaction layer, the invariant part ($h_{i,k00}$) of the atom-wise feature is passed through a linear layer and pooled to produce an energy contribution. Only the final feature is passed through an MLP to compute its energy contribution. Forces are calculated by taking derivatives of the total energy.
\Cref{tab:notations} summarizes the notation and symbols.

\begin{figure}[htb]
  \centering
  \includegraphics[width=0.9\linewidth]{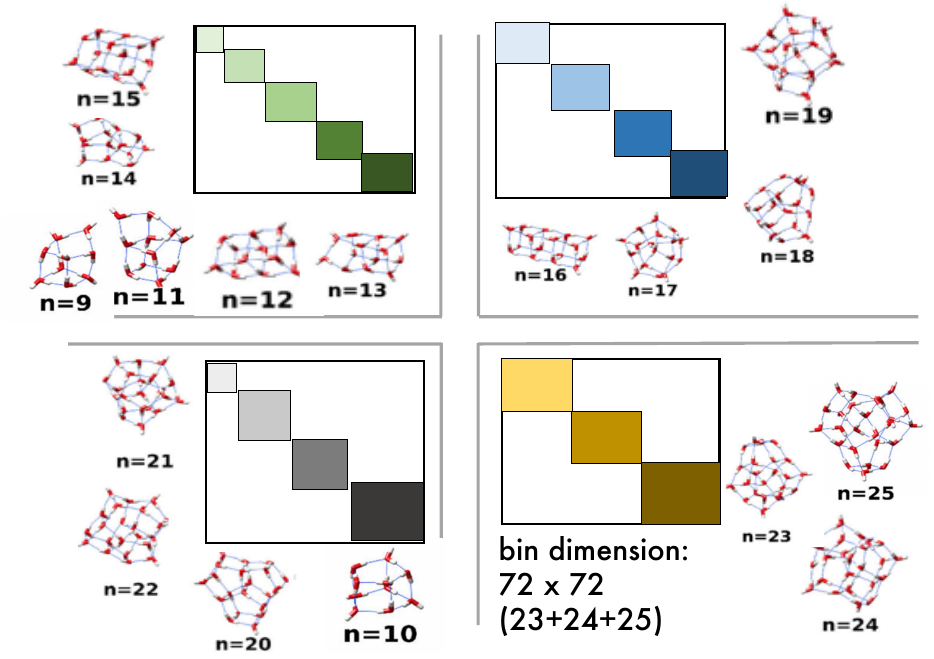}
  \vspace{-1em}
  \caption{\small Minibatch creation and distribution process of molecular graphs for a GNN model training across 4 compute nodes.}\label{fig:minibatches_bin}
\end{figure}
\section{Optimizations for Data Distribution in CFMs}\label{sec:bin_packing_load_balancer}

\textbf{Challenges in distributing molecular graphs for equivariant GNN training.}
A Chemistry foundation model is typically trained on a large set of molecular graphs drawn from various chemical systems. These graphs are generally small but diverse in terms of the vertex counts and the edge counts (\Cref{table:dataset} and \Cref{fig:dataset}).
CFMs implemented with popular graph neural network libraries such as PyTorch Geometric (PyG)~\cite{Fey_Lenssen_2019} and Distributed Deep Graph Library (DGL)~\cite{zheng2020distdgl} 
often combine a fixed number of these graphs into a single graph to create \textit{mini-batches} to process the graphs simultaneously. This is done by placing the adjacency matrices of the individual molecular graphs along the diagonal, creating a large graph with isolated subgraphs 
(for example, the colored blocks placed within each square in \Cref{fig:minibatches_bin}). Once these mini-batches are created, they are distributed across multiple GPUs to train a model.

Popular frameworks like PyG employ \textit{fixed-size} mini-batching i.e. \textit{fixed number of graphs combined in a mini-batch} (for example 5 graphs per minibatch irrespective of their sizes). However, a straightforward approach of randomly combining graphs have limitations. 
\vspace{-0.5em}
\begin{tcolorbox}[colback=bg,left=1ex,top=1ex,boxsep=0ex,bottom=1ex,width=\linewidth]
  \rulegen{}:
  Using a fixed number of molecular graphs in each mini-batch ignores the diversity in graph sizes that can lead to uneven sizes in mini-batches, impacting performance. 
Based on the sparsity pattern in each graph, the total amount of work in each mini-batch can also vary significantly. 
Consequently, when the workload is distributed across multiple GPUs, the GPU execution time can differ, with the slowest GPU (straggler) limiting overall performance.
\end{tcolorbox}

\newcommand{\myAlgFontA}{\small}%
\newcommand{\myAlgFontB}{\scriptsize}
\newcommand\mycommfont[1]{\footnotesize\textcolor{blue}{#1}}
\SetCommentSty{mycommfont}

\newcommand{\loadbalacerAlg}{
{
\begin{algorithm}[t]
\small
\caption{\textsc{Create-Balanced-Batches}}\label{algo:balanced_batches}
\KwIn{$L = (l_1, \ldots, l_N)$: list of graph vertex counts, $C$: bin capacity, $G$: number of GPUs}
\KwOut{$\mathcal{B}$: set of bins with graph indices}

$(L, I) \leftarrow \text{StableSort}(L, \text{descending} = \text{True})$ \tcp*{$I$ is the index mapping} \label{line:stablesort}
$S \leftarrow \sum_{i=1}^N l_i$\;
$M \leftarrow \left\lceil\frac{S}{C}\right\rceil$\;
$M \leftarrow \left\lceil\frac{M}{G}\right\rceil \cdot G$ 

$\mathcal{B} \leftarrow \{B_1, \ldots, B_M\}$, where $\forall j, B_j = \emptyset$ and $c(B_j) = C$\;
$p \leftarrow 1$ \tcp*{Pointer to current item in $L$}

\While{$p \leq N$ \textbf{and} $\mathcal{B} \neq \emptyset$}{
    $\mathcal{B} \leftarrow \text{StableSort}(\mathcal{B}, \text{key} = c, \text{descending} = \text{True})$\; \label{line:end_coll}
    \For{$j \leftarrow 1$ \KwTo $|\mathcal{B}|$}{\label{line:iterate}
         \If{$c(B_j) \geq l_p$}{\label{line:capacity_check}
            $B_j \leftarrow B_j \cup \{I_p\}$\;\label{line:index_add}
            $c(B_j) \leftarrow c(B_j) - l_p$\;\label{line:adjust_cap}
            $p \leftarrow p + 1$\; \label{line:pointer_adjust}
            \If{$p > N$}{
                \textbf{break}\;
            }
        }
        \Else{
            Mark $B_j$ as full\;\label{line:mark_full}
        }
    }
    $\mathcal{B}_{\text{full}} \leftarrow \{B_j \in \mathcal{B} : B_j \text{ is full}\}$\;\label{line:start_coll}
    $\mathcal{B} \leftarrow \mathcal{B} \setminus \mathcal{B}_{\text{full}}$\;

    \If{any non-full bin has less remaining capacity than a full bin}{ \label{line:second_chance_start}
        Unmark all full bins in $\mathcal{B}_{\text{full}}$\;
        $\mathcal{B} \leftarrow \mathcal{B} \cup \mathcal{B}_{\text{full}}$\;
    } \label{line:second_chance_end}
}
\If{$p \leq N$}{\label{line:call_recursive_begin}
    $\mathcal{B}_{\text{remaining}} \leftarrow \text{CREATE-BALANCED-BATCHES}((l_p, \ldots, l_N), C, G)$\; 
    $\mathcal{B} \leftarrow \mathcal{B} \cup \mathcal{B}_{\text{remaining}}$\;
}\label{line:call_recursive_end}
\Return{$\mathcal{B}$}
\end{algorithm}
}
}

\subsection{Mini-batches Creation as the Multi-objective Bin Packing Problem}

We address the challenge of efficient distribution of molecular graphs by formulating it as a multi-objective bin packing problem~\cite{poothokaran2010study,Korte2012}. Given a list of graph vertex counts and a maximum bin capacity, our goal is to create balanced batches that minimize both load imbalance and memory padding overhead, while ensuring the number of batches is a multiple of available GPUs for efficient distributed training. We develop an efficient iterative algorithm to address this challenge.
This problem can be approached as either a scheduling problem with a fixed number of bins or a bin packing problem where the number of bins is determined by the algorithm~\cite{bin_packing_multi}. 
As an example, the distribution scheme in ~\Cref{fig:minibatches_bin} employs four bins (represented with large squares) to store the molecular graphs, with the bins containing 5, 4, 3, and 4 molecular graphs in clockwise order. The bottom-right bin contains three molecular graphs with 23, 24, and 25 vertices, yielding a total \textit{token count} (combined vertex count) of 72 (dimension of the bin).

\subsubsection{Problem formulation}
Let us denote the number of vertices in a graph with vertex set $V$ by $|V|$.
We assume there are $N$ graphs to be distributed across the mini-batches (bins). The maximum number of possible bins is $N$, and we define binary variables 
$a_j$ 
to indicate 
whether bin $j$ is used ($a_j = 1$) or not ($a_j = 0$). 
This variable $a_j$ is subject to minimization. The assignment variable $b_{ij}$ indicates whether the $i$-th graph is placed in bin $j$ ($b_{ij} = 1$), otherwise $b_{ij} = 0$. The capacity of each bin i.e. the maximum number of vertices is denoted by $C$.
The width and the height of each (square) bin is $W$(=$C$). 
We want to find a $t \in \mathbb{N}$ and assignment $f: \{1, \cdots, N\} \rightarrow\{1, \cdots t\}$ with objective functions


\begin{equation}\label{eqn:min_bin}
\begin{aligned}
\min \quad \sum_{j=1}^{N}{a_{j}} 
\end{aligned}
\end{equation}
\begin{equation}\label{eqn:min_memory_padding}
\begin{aligned}
\text{min} \sum_{j=1}^{t} \left[\sum_{i=1}^{N} \frac{b_{ij} |V_i|^2}{W^2}\right]   
\end{aligned}
\end{equation}
\begin{equation}\label{eqn:load_balance}
\begin{aligned}
\text{min} \max_{j,k} \left|\sum_{i=1}^{N} (b_{ij} |V_i|^2) - \sum_{i=1}^{N} (b_{ik} |V_i|^2)\right|   
\end{aligned}
\end{equation}
%
\begin{equation}\label{eqn:capacity}
\begin{aligned}
\sum_{i=1}^N |V_i|b_{ij}\leq C a_j \quad  \forall j.
    \end{aligned}
 \end{equation}  
\begin{equation}\label{eqn:assignment}
    \begin{aligned}
        \textrm{s.t.} \quad & \sum_{j=1}^{t}  b_{ij} =1\quad   \forall i \\ 
    \end{aligned}
\end{equation}

\cref{eqn:min_bin,eqn:min_memory_padding,eqn:load_balance} states the following objectives: we aim to minimize the number of bins, 
while minimizing zero-padding memory in each bin (mini-batch), 
and the difference between the amount of space filled by molecular graphs in any two bins should be minimal. \Cref{eqn:capacity} is the capacity constraint. The assignment constraint (\Cref{,eqn:assignment}) ensures that each graph is assigned to a bin. Given a fixed bin capacity, padding refers to the unused memory that must be zero-filled when no remaining graph fits in the available space.

\loadbalacerAlg{}
\subsection{Iterative Algorithm for Multi-Objective Bin Packing}
The multi-objective bin packing for molecular graphs is a NP-hard problem~\cite{GareyJohnson1979}. To  address the multi-objective bin packing problem, we propose an iterative algorithm (\Cref{algo:balanced_batches}) that balances computational load while minimizing padding overhead. 
%
%
The algorithm takes as input a set of graphs with associated vertex counts, a maximum capacity per bin, and the number of GPUs used. The objective is to distribute these graphs across $M$ bins, a multiple of the number of GPUs, in a way that balances the load across GPUs while respecting capacity constraints.

Initially, graphs are sorted by their vertex count in descending order to handle larger graphs first (\Lineref{line:stablesort}). In each iteration, bins are stably sorted by their remaining capacity to prioritize fuller bins (\Lineref{line:end_coll}). 
Within the iteration, the algorithm steps through all (remaining) bins once (\Lineref{line:iterate}). The algorithm iteratively assigns a graph to each bin. If a bin's remaining capacity is higher than the size of the graph currently considered (\Lineref{line:capacity_check}), the corresponding index is appended to the bin (\Lineref{line:index_add}), the remaining capacity reduced (\Lineref{line:adjust_cap}), and the pointer advances (\Lineref{line:pointer_adjust}). 
When a bin cannot accommodate the current item, it is marked as full.(\Lineref{line:mark_full}). At the end of the iteration, the full bins are collected (\Lineref{line:start_coll}). Intuitively, 
the process can be thought of as the cyclic distribution of the sorted molecular graphs  across the available bins. 

However, the algorithm also incorporates an adaptive bin management mechanism: when a bin marked as ``not full'' achieves a higher occupancy than bins previously marked as ``full'', it indicates we have succeeded in packing more items than was deemed possible earlier. At this point, all bins previously marked as 'full' are reconsidered and returned to the active pool (\linesref{line:second_chance_start}{line:second_chance_end}), as they may now be able to accommodate the currently smaller remaining graphs. This opportunistic reassessment of bin availability helps achieve higher packing density across all bins.

This process continues until all graphs are assigned or no suitable bins remain. If any graphs are left unassigned, the process is repeated for remaining graphs (\linesref{line:call_recursive_begin}{line:call_recursive_end}). Stable sorting is used to maintain consistency across processes. The algorithm produces locally uniform batches, resulting in balanced load across GPUs. 

Note that, our algorithm differs from well-known best-fit and first-fit algorithms~\cite{GareyJohnson1979} for bin packing in several ways. First, methods like best-fit algorithm aims to minimize waste in each individual bin placement. In contrast, our proposed algorithm optimizes for both bin utilization and load balance across adjacent bins/GPUs. Our algorithm distributes the graphs among the bins, compared to the best-fit algorithm. The later fills in a bin until full and then opens a new bin. Additionally, our algorithm pre-processes all molecular graphs by sorting them by vertex counts and has a global view of the workload optimization.

\subsubsection{Implementation details}
We have modified PyTorch's \texttt{Distributed Sampler} to implement our algorithm. The sampler is edited to become a batch sampler, where at the beginning of every epoch, the batches are determined with Algorithm \ref{algo:balanced_batches}. Note that the ``size'' of a graph can be defined arbitrarily, in addition to above-mentioned vertex count, such as edge count, or a function of both.

\subsubsection{Time complexity}
The initial sorting of graphs takes $O(N \log N)$, where $N$ is the total number of graphs. In each iteration, sorting the remaining capacity of bins takes $O(M \log M)$, where $M$ is the number of bins. The number of iterations is $O(\frac{N}{M})$. Therefore the overall time complexity of the algorithm is $O(N \log N) + O(N \log M)$. On a single CPU, with $\sim$1M molecular graph samples and $\sim$100k batches, preparing the batches with this algorithm takes about one second.

\begin{figure*}[htb]
  \centering
  \includegraphics[width=0.9\linewidth]{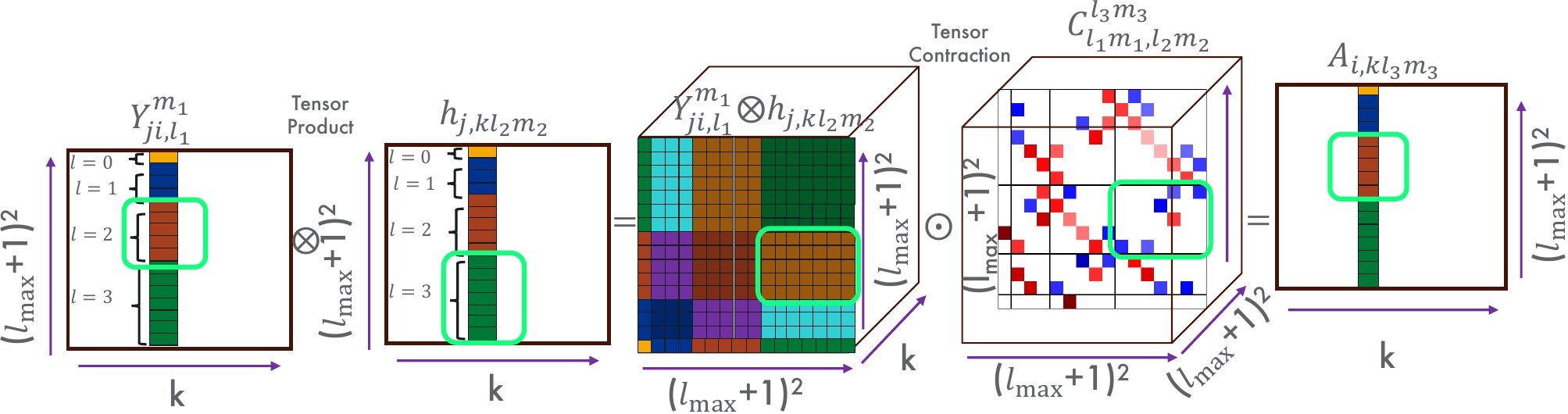}
  \vspace{-1.2em}
  \caption{\small A schematic diagram of the tensor contraction performed in Algorithm \ref{alg:channel_wise}. Conceptually, two equivariant tensors, $Y_{ji,l_1}^{m_1}$ and $h_{j,kl_2m_2}$, are multiplied via the normal tensor product to form the product tensor $Y_{ji,l_1}^{m_1}\otimes h_{l_2m_2}$ which is contracted using the Clebsch-Gordon coefficients to form the output equivariant tensor $A_{i,kl_3m_3}$. The additional channel dimension $k$ is mixed by the radial embedding $R_{ji,kl_1l_2l_3}$ (not shown). In practice the product is computed for every combination of $l_1,l_2,l_3$ up to $l_{\mathrm{max}}$. The green boxes depict the $l_1=2,l_2=3,l_3=2$ selection and the Clebsch-Gordon coefficient matrix is shown for $l_1=2,l_2=3,l_3=2$ and $m_3=-1$, which is highly sparse.
  }
  \label{fig:contraction}
\end{figure*}

\section{ Computational Kernel Optimization  
}\label{sec:sym_cont_channeL_prod}
We now focus on the most computationally expensive operations  within the MACE architecture's interaction layer and the product block (\Cref{fig:model_architecture_symmtric_contraction} c-d): the symmetric tensor contraction and the tensor product to construct the messages, 
referred to as the channelwise tensor product (TP). Tensor contraction serves as the fundamental operation in MACE~\cite{batatia2022mace} and other prominent models such as NequIP~\cite{batzner20223}, Allegro~\cite{musaelian2023learning} and Cormorant~\cite{anderson2019cormorant}.

\subsection{Symmetric Tensor Contraction: Optimization Needs}

\begin{algorithm}
\small
\SetAlgoLined
\KwIn{$C^{l_3m_3}_{l_1m_1,l_2m_2}$: Clebsch-Gordan coefficients}
\KwIn{$R^{(t)}_{ji,kl_1l_2l_3}$: Precomputed weights for edge $ji$}
\KwIn{$h^{(t)}_{j,klm}$: Features of atom $j$}
\KwIn{$Y^{m}_{ji,l}$: Spherical harmonics for edge $ji$}
\KwIn{$N_\text{edges}$: Number of edges}
\KwIn{$N_\text{channels}$: Number of channels}
\KwIn{$L_\text{max}$: Maximum irreps order}

\For{$ji = 1$ \KwTo $N_\text{edges}$}{
    \For{$k = 1$ \KwTo $N_\text{channels}$}{
        \For{$l_3 = 0$ \KwTo $L_\text{max}$}{
            $A^{(t)}_{ji,kl_3m_3} \leftarrow 0$\;
            \For{all combinations of $l_1m_1$, $l_2m_2$}{
                $A^{(t)}_{ji,kl_3m_3} \leftarrow A^{(t)}_{ji,kl_3m_3} + 
                C^{l_3m_3}_{l_1m_1,l_2m_2}
                R^{(t)}_{ji,kl_1l_2l_3} 
                Y^{m_1}_{ji,l_1}
                h^{(t)}_{j,kl_2m_2}$\;\label{line:tp}
            }
        }
    }
}
\KwOut{$A^{(t)}_{ji,kl_3m_3}$}
\caption{Construction of original messages in MACE}\label{alg:channel_wise}
\end{algorithm}
%


\subsubsection{Bottleneck in atomic basis ($A$) construction (\Cref{alg:channel_wise}).} In MACE, messages are constructed from vertex features and the relative vertex positions (the 
 interaction block in \Cref{fig:model_architecture_symmtric_contraction} (c)).  The first step to build messages is the channelwise TP (\Cref{fig:contraction}) presented in \Cref{alg:channel_wise}, Line 6. Let us consider the message from atom $j$ to atom $i$ at epoch $t$: the first factor entering in the message are vertex features $h^{(t)}_{j,klm}$, which represent properties of the sender atom (vertex) in channel $k$ and of order $l$. The second factor is the relative position of atoms $i$ and $j$ expressed in terms of spherical harmonics $Y^m_l(\hat{r}_{ji})=Y^m_{ji,l}$. The third factor, the learnable weights, $R^{(t)}_{ji,kl_1l_2l_3}$, are pre-computed from a multilayer perceptron taking a radial embedding of the interatomic distances $r_{ji}$ as input, and determine how much importance is given to different types of interaction. This algorithm performs a convolution-like operation: it combines information from the vertex features $h^{(t)}_{j,klm}$ and the geometric relationship between vertices $Y^m_{ji,l}$ using learnable weights and the Clebsch-Gordan coefficients, $C^{l_3m_3}_{l_1m_1,l_2m_2}$ (Lines 5-6). 
 
 This step is particularly expensive because the product is computed for each edge of the graph. The parameter $L_\text{max}$ limits the maximum order of the resulting message and can be tuned to obtain a higher order (but more expensive) description.
 


\textbf{Unexploited sparsity in Clebsch-Gordan coefficient access.} In \Cref{alg:channel_wise}, when constructing messages via Clebsch-Gordan coefficients $C^{l_3m_3}_{l_1m_1,l_2m_2}$, only specific combinations of $(l1,m1,l2,m2)$ yield non-zero coefficients. The number of non-zero elements typically represents less than 20\% of possible coefficient combinations. 
This sparsity pattern is deterministic and known at compile time. 
Current approach does not take CG tensor's sparsity into consideration and applies dense matrix multiplication. 
\vspace{-0.5em}
\begin{tcolorbox}[colback=bg,left=1ex,top=1ex,boxsep=0ex,bottom=1ex,width=\linewidth]
  \rulegen{}:
  The existing approach doesn't consider the sparsity 
  of Clebsch-Gordan coefficients, leading to unnecessary computations in dense matrix multiplication, while exploiting these properties could significantly reduce storage and computational requirements.
\end{tcolorbox}
%

\begin{algorithm}
\small
\SetAlgoLined
\KwIn{$A^{(t)}_{i,klm}$: Atomic basis features}
\KwIn{$C^{LM}_{\mathbf{lm}}$: Generalized Clebsch-Gordan coefficients}
\KwIn{$W^{(t)}_{zkL,\mathbf{l}}$: Learnable weights}
\KwIn{$N_\text{atoms}$: Number of atoms}
\KwIn{$N_\text{channels}$: Number of channels}
\KwIn{$L_\text{max}$: Maximum irreps order}
\KwIn{$\nu_\text{max}$: Maximum correlation order}

\For{$i = 1$ \KwTo $N_\text{atoms}$}{
    \For{$k = 1$ \KwTo $N_\text{channels}$}{
        \For{$L = 0$ \KwTo $L_\text{max}$}{
            $m^{(t)}_{i,kLM} \leftarrow 0$\;
            \For{$\nu = 1$ \KwTo $\nu_\text{max}$}{
                \For{all combinations $\mathbf{lm}$ of $l_1m_1, \ldots, l_{\nu}m_{\nu}$}{
                    $\text{prod} \leftarrow 1$\;
                    \For{$\xi = 1$ \KwTo $\nu$}{
                        $\text{prod} \leftarrow \text{prod} \cdot  A^{(t)}_{i,k l_\xi m_\xi}$\;
                    }
                    $m^{(t)}_{i,kLM} \leftarrow m^{(t)}_{i,kLM} + W^{(t)}_{z_i kL,\mathbf{l}} C^{LM}_{\mathbf{lm}} \cdot \text{prod}$\;
                }
            }
        }
    }
}

\KwOut{$m^{(t)}_{i,kLM}$}
\caption{Higher-order feature computation in MACE}\label{alg:mace_contraction}
\end{algorithm}

\vspace{-0.5em}
\subsubsection{Bottleneck in message construction i.e. ``pooling'' (\Cref{alg:mace_contraction}).} After messages from all neighbors $j$ are received, they are summed over $j$ (by channel $k$ and order $l$) on the receiving vertex $i$. After a further linear combination (between terms $k$ of the same order), the central operation of MACE can be performed: the symmetric tensor contraction (Line 10 in \Cref{alg:mace_contraction}). In this operation, multiple copies of the message are contracted (through a tensor product) in order to produce features of higher order (\Cref{fig:model_architecture_symmtric_contraction} (d)).

As detailed in \Cref{alg:mace_contraction}, a product of up to $\nu$ copies of the messages $A^{(t)}_{i,klm}$ are computed, for a given channel $k$ and considering all possible combinations of $l_1m_1, \ldots, l_{\nu}m_{\nu}$ that would result in a nonzero contribution to a given $LM$, up to $L_\text{max}$ (Lines 5-9).
Similar to \Cref{alg:channel_wise}, the high sparsity of the $C^{LM}_{\mathbf{lm}}$ tensor makes this symmetric tensor contraction particularly attractive for optimization.


\textbf{Inefficiency in symmetric tensor contraction implementation in existing libraries.} Tensor contractions, specifically symmetric tensor contractions in the MACE model are originally implemented via contracting multiple tensor segments as a chain of small kernels in auto-differentiation deep learning frameworks such as \texttt{PyTorch-based e3nn~\cite{e3nn_software}}. 
For tensor contractions, features for a tuple $(l, m)$ correspond to tensor segments which are then naively contracted for each $(l_1, l_2, l_3)$ or even for each $(l_1, m_1, l_2, m_2, l_3, m_3)$ in separate kernel calls. This results in inefficiency in modern GPUs.
Naively chaining these primitives thus results in repeated loads and stores of intermediate results into the global memory.
\vspace{-1em}
\begin{tcolorbox}[colback=bg,left=1ex,top=1ex,boxsep=0ex,bottom=1ex,width=\linewidth]
  \rulegen{}:
  Existing implementations in frameworks like PyTorch e3nn perform symmetric tensor contractions by breaking them into many small separate kernel calls for each combination of quantum numbers (l,m), leading to excessive global memory access, poor GPU utilization, and frequent small kernel launch overhead.
\end{tcolorbox}

\subsection{
Optimization Strategies}
To address the inefficient symmetric tensor contractions and unexploited Clebsch-Gordan coefficient sparsity patterns identified above in  \texttt{Pytorch e3nn} library, that provides the building blocks for many well-known equivariant GNNs including MACE, and to improve the overall performance of \Cref{alg:channel_wise} and \Cref{alg:mace_contraction}, we apply the following optimization techniques. Due to space constraint, we only provide highlighted optimizations in code for \Cref{alg:mace_contraction} in \Cref{code:opt_alg}.  
\subsubsection{Applying kernel fusion} To avoid the problem of launching multiple small kernels, \textit{kernel fusion}
combines as many of these operations in a single GPU kernel thus allowing for keeping intermediate results in local memory as long as possible. Local memory is faster by several orders of magnitude, thus improving the overall  performance. 
We adopt the kernel fusion technique for the symmetric tensor contraction, replacing many natively-supported operations with a single special-purpose GPU kernel. Kernel fusion executes the entirety of \Cref{alg:channel_wise} and \Cref{alg:mace_contraction} in a single kernel with parallelism derived from their outermost loops. By applying kernel fusion technique to combine multiple small kernels in a larger GPU kernel improves kernel launch time, enables compilers and machine learning frameworks to automatically detect and fuse kernels and operators.

\begin{listing}
\caption{Optimized \Cref{alg:mace_contraction}}
\label{code:opt_alg}
\setminted{fontsize=\footnotesize}
\begin{tcolorbox}[colback=bg,left=1ex,top=-1ex,boxsep=0ex,bottom=-1ex,width=1.05\linewidth]
\begin{minted}[escapeinside=||]{c++}
__global__ void optimized_higher_order_features(
 const float* A_t, const float* CG_LM, const float* W_t,
 float* m_t,  const int* z_i, const int I, const int K,
 const int L_max, const int M_max, const int nu_max) {
 // Shared memory for input features and intermediate results
 __shared__ float s_A[BLOCK_SIZE][MAX_L][MAX_M];  |\codecircle{1}|  //Input features
 __shared__ float s_B[MAX_NU][MAX_ETA][MAX_K];//Intermediate res.
    
 const int tid = threadIdx.x;
 const int wid = tid / WARP_SIZE; const int lid = tid % WARP_SIZE;
 const int atom_idx = blockIdx.x;  // Each block handles one atom
 const int L = blockIdx.y;  const int M = blockIdx.z;  

 //Load atomic features into shared memory(vectorized and coalesced)
 for(int k = wid; k < K; k += gridDim.x) {
  float4 a_vals = *reinterpret_cast<const float4*>(
   &A_t[atom_idx * K * L_max * M_max + k * L_max * M_max]);  |\codecircle{2}|
        
  #pragma unroll 4
  for(int idx = 0; idx < 4; idx++) {
   int l = (lid * 4 + idx) / M_max;
   int m = (lid * 4 + idx) % M_max;
   if(l < L_max && m < 2*l + 1) {
    s_A[tid][l][m] = ((float*)&a_vals)[idx];
   }}}
   __syncthreads();

  // Process each correlation order
  float result = 0.0f;
  for(int nu = 1; nu <= nu_max; nu++) {
   // Each warp processes different combination of indices
   for(int eta=wid;eta<num_coupling_patterns[nu];eta +=gridDim.x) {
    float prod = 1.0f;
    // Process tensor products within warp
    #pragma unroll
    for(int xi = 0; xi < nu; xi++) {
     // Get indices for this correlation order
     int l_idx = get_l_index(nu, eta, xi);
     int m_idx = lid;  // Distribute m indices across warp
     if(m_idx < 2*l_idx + 1) {
      // Load and multiply feature values
      float a_val = s_A[tid][l_idx][m_idx];
      // Warp-wide product using shuffle
      |\codecircle{5}|  #pragma unroll
      for(int offset = WARP_SIZE/2; offset > 0; offset /= 2) {
       float other = __shfl_xor_sync(0xffffffff, a_val, offset);  |\codecircle{3}|
       a_val *= other; }
      // First thread in warp accumulates
      if(lid == 0) prod *= a_val;       
       }}
       // Get corresponding CG coefficient sparse index
      float cg = CG_LM[get_cg_index(nu, eta, L, M)];  |\codecircle{4}|
      int z = z_i[atom_idx];
      float weight = W_t[get_weight_index(z, K, L, eta)];
      // Accumulate contribution
      if(lid == 0) {  // First thread in warp
       float contrib = cg * weight * prod;
       // Atomic add to handle concurrent updates
       atomicAdd(&m_t[get_output_index(atom_idx,K,L,M)], contrib);
  }}}}
\end{minted}
\end{tcolorbox}
\end{listing}


However, there are several challenges associated with implementing kernel fusion. Specifically, kernel fusion requires careful balancing of resource usage while extracting performance. The main challenge is resource exhaustion and reduced parallelism. While fused kernels require more resources (registers, shared memory, instruction cache) than the unfused ones, we mitigate this through techniques like reusing the shared memory buffers (\circleb{1} in \Cref{code:opt_alg}). Our molecular graph domain provides sufficient parallelism through the number of edges and vertices to maintain efficient execution despite these resource constraints.

\subsubsection{Exploiting Clebsch-Gordan (CG) tensor sparsity} 
According to the law in quantum mechanics, there is a set of key rules that determine which combinations give non-zero CG coefficients. Specifically, for $C^{l_3m_3}_{l_1m_1,l_2m_2}$, the $l$ values must satisfy the ``triangle rule'': $|l_1 - l_2| \leq l_3 \leq l_1 + l_2$. The $m$ values must satisfy: $m_1 + m_2 = m_3$, $|m_1| \leq l_1$, $|m_2| \leq l_2$, and $|m_3| \leq l_3$. Since these rules are known at compile time, we can pre-compute all valid combinations,  store only non-zero coefficients, and create lookup tables for fast access. Additionally, CG-tensor sparsity optimization converts operations on the \texttt{CG} tensor into sparse multiplications focusing only on non-zero elements (\circleb{4} in \Cref{code:opt_alg}).

\vspace{-0.5em}
\subsubsection{Maximizing bandwidth utilization} Memory bandwidth is optimized through efficient thread-layouts for coalesced access patterns and vectorized load/store operations. In particular, for \Cref{alg:channel_wise}, the implementation can utilize vectorized loads through \texttt{float4} operations to efficiently load \textit{blocks} of the input tensors $R$ and $h$ and coalesced memory access so that adjacent threads access adjacent memory locations. Similarly, in \Cref{alg:mace_contraction},  the input features $A$ can be loaded using vectorized operations and stored in shared memory (\circleb{2} in \Cref{code:opt_alg}).  In addition, shared memory can be utilized by storing frequently accessed data such as the Clebsch-Gordan coefficients and spherical harmonics $Y$.  Shared memory utilization reduces global memory access by preloading and operating on data in local memory.

\vspace{-0.5em}
\subsubsection{Utilizing warp-level primitives and loop unrolling} For \Cref{alg:channel_wise}, the computation is organized such that each thread block handles a specific $(l3,m3)$ combination, with warps collaboratively processing different $(l1,m1,l2,m2)$ combinations. Within each warp, threads exploit the regular structure of the indices by distributing the $m_2$ indices across threads while sharing the same $l_2$ value, enabling efficient parallel reduction using warp shuffle. 

For \Cref{alg:mace_contraction}, the implementation assigns each thread block to process a specific atom index $i$ and output indices $(L,M)$, with warps handling different coupling patterns $\eta$ for each correlation order $\nu$. The computation can exploit the structure of higher-order correlations by using warp-level butterfly exchange patterns through \texttt{\_\_shfl\_xor\_sync} (\circleb{3} in \Cref{code:opt_alg}) operations to compute tensor products efficiently. In addition, loop unrolling (for example \circleb{5} in \Cref{code:opt_alg})  eliminate loop overhead in warp-level operations where the number of iterations is known at compile time, allowing better instruction scheduling and register allocation by the compiler.

\vspace{-0.5em}
\section{Evaluation}\label{sec:eval}

\begin{figure*}[htb]
  \vspace*{-1.5em}
  \centering
  \includegraphics[width=0.7\linewidth]{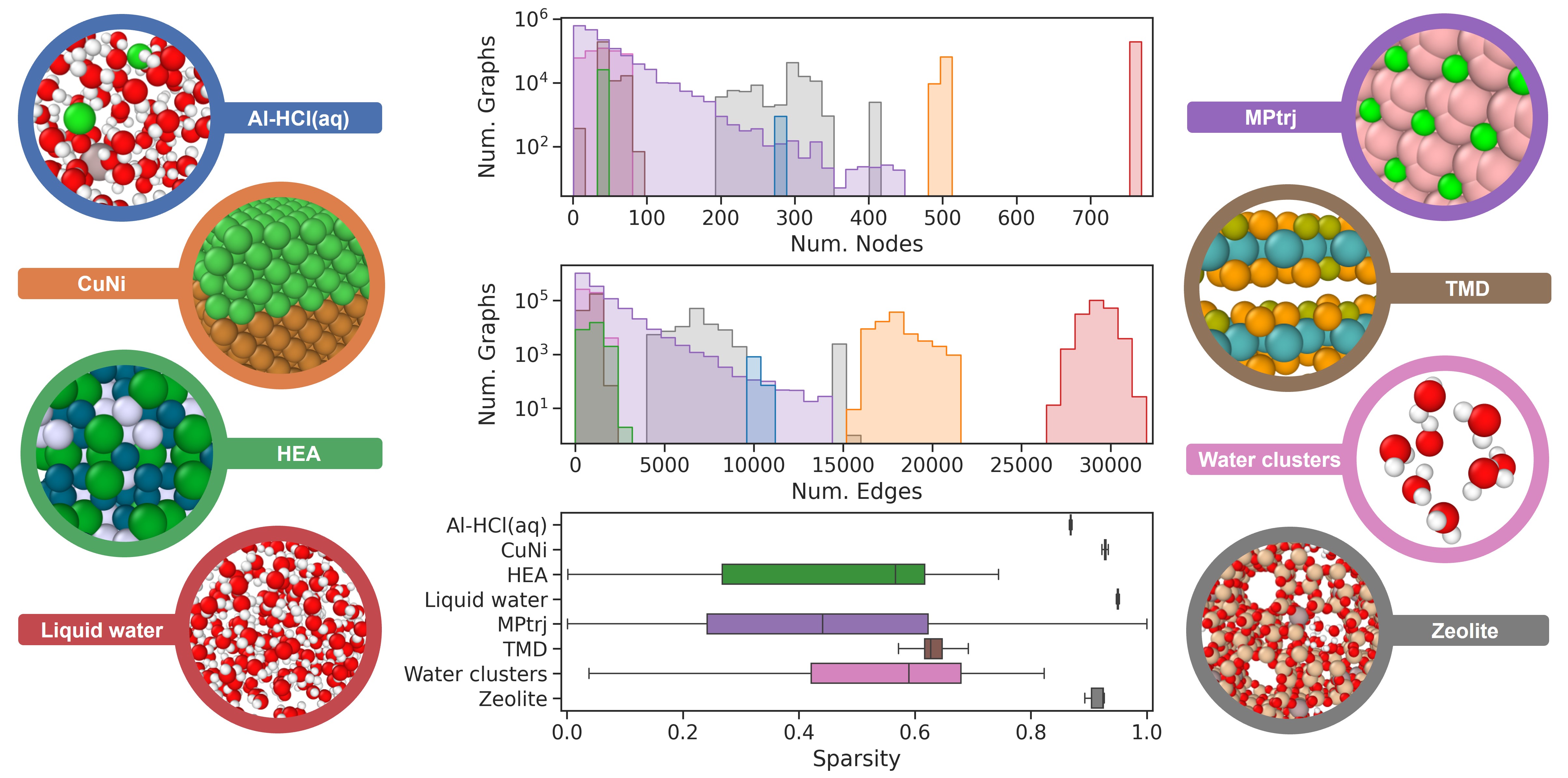}
  \vspace{-1.5em}
  \caption{\small Information on the different chemical systems included in our combined graph dataset: histograms of the vertex and edge counts (in log scale) and sparsity distributions when $r_{\mathrm{cutoff}}=4.5$ \r{A}.}
  \label{fig:dataset}
  \vspace*{-1.5em}
\end{figure*}

We investigate the following research questions to evaluate the effectiveness of our proposed optimizations:

\begin{itemize}[noitemsep,topsep=0pt,leftmargin=*]
    \item \textbf{Effectiveness:} 
    How do our proposed load balancing and kernel optimizations affect model training? We conduct comprehensive ablation studies (\Cref{sec:ablation}) and scaling experiments (\Cref{subsec:scaling}) to quantify the individual and combined effects of these optimizations. 
    
    \item \textbf{Robustness:} 
    How do our optimizations perform across datasets with varying size and sparsity? We evaluate performance across diverse chemical systems ranging from small molecules to large periodic structures (\Cref{subsec:scaling}). 
    
    \item \textbf{Optimal Configuration:} 
    What are the optimal configuration parameters for maximum GPU utilization? We empirically determine the optimal bin capacity that balances computational efficiency with memory constraints (\Cref{sec:bin_capacity_det}). 
    
    \item \textbf{Performance Analysis:} 
    How do our optimizations affect the computation-communication balance? We answer this via profiling the application
\end{itemize}


\vspace{-1em}
\subsection{Experimental Setup}
\subsubsection{Dataset}
\label{sec:dataset}

\begin{table}[t]
\small
    \centering
    \caption{\small Size of datasets and the range of vertices in each system.}
    \vspace{-1em}
    \begin{tabular}{p{5.7em}p{4.7em}p{3.8em}p{6em}} \toprule 
        Dataset &  Num.~Graphs & Proportion & Num.~Vertices  \\ \midrule
        Al-HCl(aq)	&	884	& <1\% &	281--281	\\
        CuNi	&	74335	& 3\% &	492--500	\\
        HEA	&	25628	& 1\% &	36--48	\\
        Liquid water	 &	190267 & 7\%	&	768--768	\\
        MPtrj	&	1580312	& 60\% &	1--444	\\
        TMD	&	219627	& 8\% &	16--96	\\
        Water clusters	 &	460000 & 17\%	&	9--75	\\
        Zeolite	&	99770	& 4\% &	203--408	\\
        \bottomrule
    \end{tabular}
    \label{table:dataset}
\end{table}

We combine eight datasets with different chemical systems 
to create a 2.65M graph dataset. Most notable about this collection is the large variation in the number of atoms in each system, which ranges from 1 to 768. When defining molecular graphs, each vertex represents an atom and edges are placed between atom pairs within a radius $r_{\mathrm{cutoff}}$ of 4 \r{A} (including distances across periodic boundary conditions). Despite the common $r_{\mathrm{cutoff}}$, the sparsity profiles are highly diverse. 

\textbf{Diversity in size.} Figure \ref{fig:dataset} and Table \ref{table:dataset} give information on the included chemical systems and their resulting molecular graphs. The Materials Project Trajectory (MPtrj) database \cite{deng2023chgnet}, which was used to train MACE-MP-0 \cite{batatia2023foundation}, comprises 60\% of the combined dataset and contains the most diverse set of systems. The second largest dataset in our combined set, comprising 17\%, contains molecular dynamics simulations of water clusters  \cite{helal2024acceleration}. A transition metal dichalcogenide (TMD) dataset comprises 8\% \cite{muller2023open}, and a liquid water dataset comprises 7\%. Notably, the liquid water samples are the largest in the combined set with all samples containing 768 atoms. The remaining datasets comprised of zeolite, copper nickle alloy (CuNi) \cite{sprueill2023active}, high entropy alloy (HEA) \cite{lopanitsyna2023modeling}, and aluminum ions in aqueous hydrochloric acid (Al-HCl(aq)), each making up <5\% of the combined. 

\textbf{Diversity in long/short-range interactions.} All systems in the liquid water and CuNi datasets are larger than the largest system in the MPtrj dataset. Notably, these systems represent different types of long-range interactions. The CuNi systems are crystalline with mostly face-centered cubic symmetry; therefore, the ordering is highly regular throughout these systems. In contrast, the liquid water systems are amorphous. Short-range order is present due to hydrogen bonding between neighboring water molecules, but no long-range symmetry is present. However, the absence of symmetry does not imply that long-range interactions are insignificant. Long-range non-bonding interactions and electrostatic effects can still have an effect on system properties.

The MPtrj, TMD, and HEA datasets contain highly ordered crystalline structures, while the water cluster and Al-HCl(aq) datasets contain only short-range ordering. The zeolite dataset has long-range ordering in the zeolite structure, with amorphous groupings of solvent molecules within the zeolite pores.


\textbf{Dataset variants.} The strong scaling experiments were performed based on the combinations mentioned in~\Cref{table:dataset}, with an approximate size of 2.65 million graphs (samples), with diverse vertex count, edge count and sparsity. For the weak scaling experiments, the combined (large) dataset in~\Cref{table:dataset} were split into two additional subsets: small (with $\approx$ 600,000 graphs), medium (with $\approx$ 1.2 million samples). 
\vspace{-0.6em}


\subsubsection{Software Configuration} 
We used PyTorch version 2.3.1 and CUDA version 12.2. The number of workers for the PyTorch \texttt{DataLoader} \texttt{num\_workers} is set to 12 for pre-processing and  data loading (found to be the optimal for our system). The \texttt{pin\_memory} functionality is enabled to leverage faster CPU to GPU data transfer by using pinned memory.  The data is distributed using PyTorch's DistributedDataParallel (DDP) module, where mini-batches of graphs are distributed across GPUs, with the model copied on each GPU. 
\vspace{-0.5em}
\subsubsection{Hardware Configuration}
The experiments were conducted on a system with 4 NVIDIA A100 GPUs/compute node, and 80 GB main memory/GPU. Each compute node also has a single AMD EPYC 7763 CPU, with 64 cores and 256 GB of DDR4 DRAM/CPU. The system interconnect is a 3-hop dragonfly topology.

\vspace{-0.5em}
\subsection{Hyperparameters}
We set the learning rate to 0.005 in our experiments. For baseline minibatching, we specified a batch size of 6 (with systems with large atom counts) to 8 (with systems with small-medium atom counts) without running out of memory and report the average per epoch runtime. We used the Adam optimizer and an exponential moving average learning scheduler, along with the \texttt{weighted} loss function. 

Each training was performed with two layers of MACE as demonstrated to be sufficient by~\cite{batatia2022mace} due to the application of higher body order to construct equivariant messages. The body order of the message was set to 4. For load balancer, we set the maximum total token size (number of atoms) in a minibatch (bin) to 3072. For training, \texttt{Float32} datatype was used. The maximum radial cut-off for atom neighborhood (which influences the graph connectivity) was set to 4.5 \AA. 

The irreducible representations of the messages was set to \texttt{128x0e + 128x1o} according to the notation of the \texttt{e3nn} library~\cite{e3nn_software} (meaning direct sum of two feature: 128 scalar features that are invariant under inversion and
128 vector features that change sign under inversion).  Highest $\ell$ of the spherical harmonics was set to 3. Both the correlation order of each layer and the max $L$ were set to 2. We use 8 Bessel functions for the radial basis.

\subsection{Ablation Study}\label{sec:ablation}

\begin{figure}[t]
  \centering
  \includegraphics[width=0.9\linewidth]{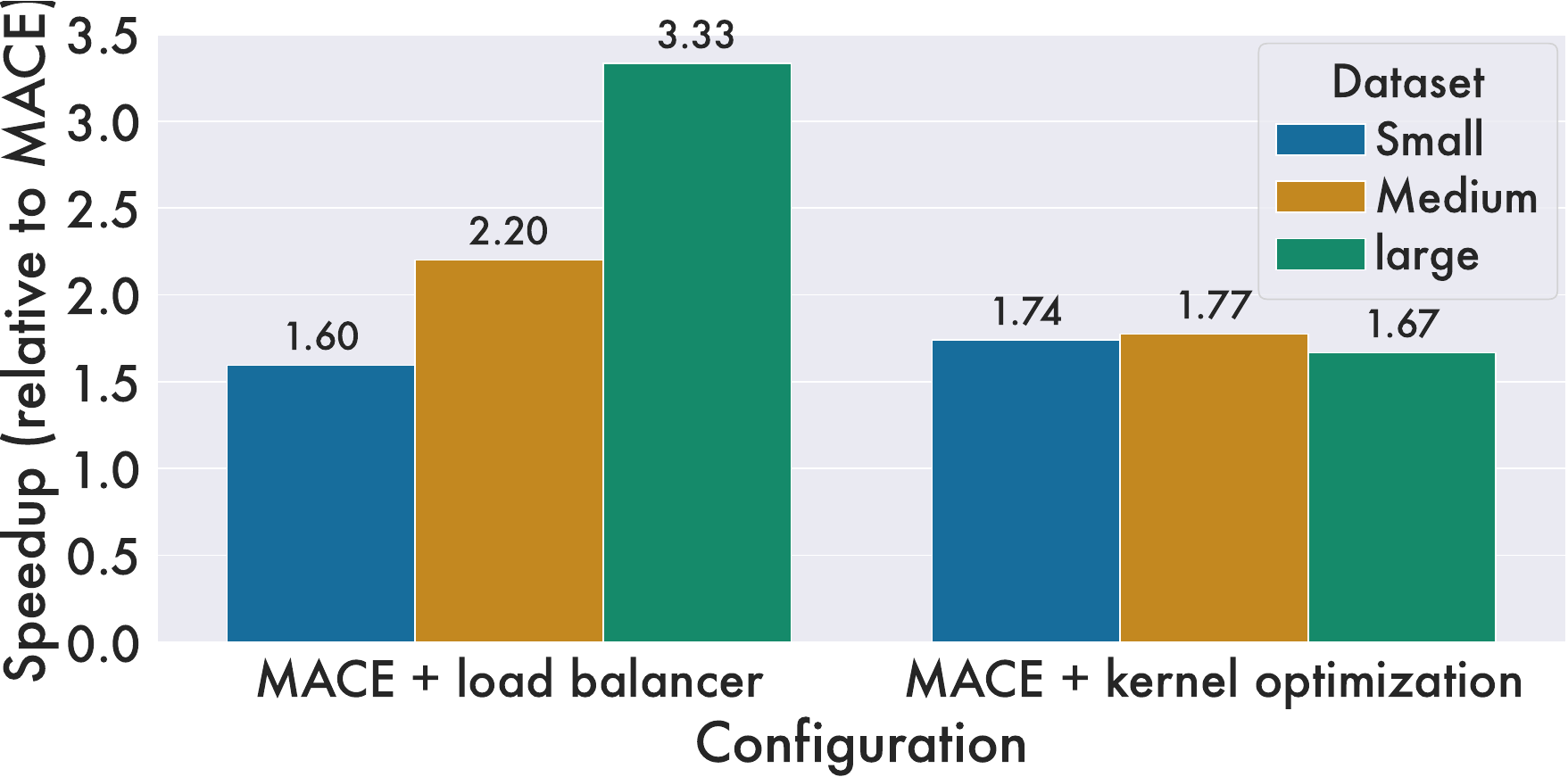}
   \vspace{-1.em}
  \caption{\small Ablation result of the proposed optimizations.}
  \label{fig:ablation}
  \vspace{0.5em}
\end{figure}

\begin{figure}[t]
  \centering
  \includegraphics[width=0.75\linewidth]{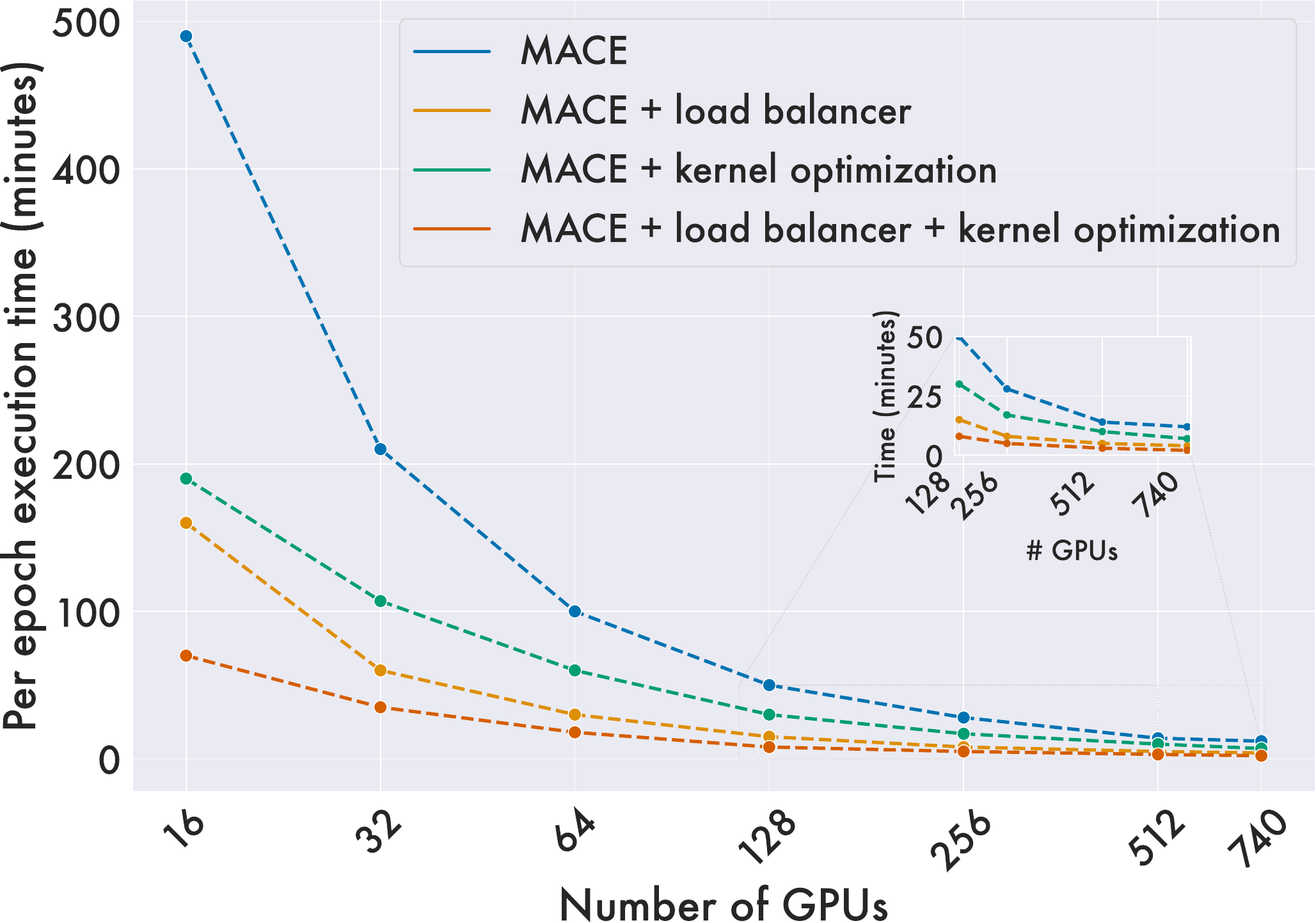}
  \vspace{-1.em}
  \caption{\small Strong scaling result with 2.6M samples.}
  \label{fig:strong_scaling}
\end{figure}

We first conducted an ablation study on the individual impact of the load balancer and the optimization of the symmetric tensor contraction on training MACE 
with different dataset sizes. For this study, we ran our experiments on 16, 32 and 64 compute nodes for small, medium and large dataset respectively. We measure the speedup of a model variation as the average per epoch execution time w.r.t. the original MACE implementation. The results are reported  in~\Cref{fig:ablation}. As can be observed from the figure, load balancing has significant impact on the performance of the model. The improvement is particularly evident with large dataset, which has the widest variation in vertex and edge counts -- up to $3.3\times$. With the kernel optimizations, training performance improvements of the model with different datasets are generally the same, up to $1.7\times$.

\subsection{Strong and Weak Scaling Results}\label{subsec:scaling}
\subsubsection{Strong Scaling}
The strong scaling performance of the MACE model, with and without our proposed optimizations is reported in~\Cref{fig:strong_scaling}, comparing the per-epoch execution time as the number of GPUs increases. The analysis was done with the combined dataset of $\approx 2.6$ million graphs. As the number of GPUs increases, all configurations show improved execution time. 

\begin{figure}[t]
  \centering
  \includegraphics[width=0.8\linewidth]{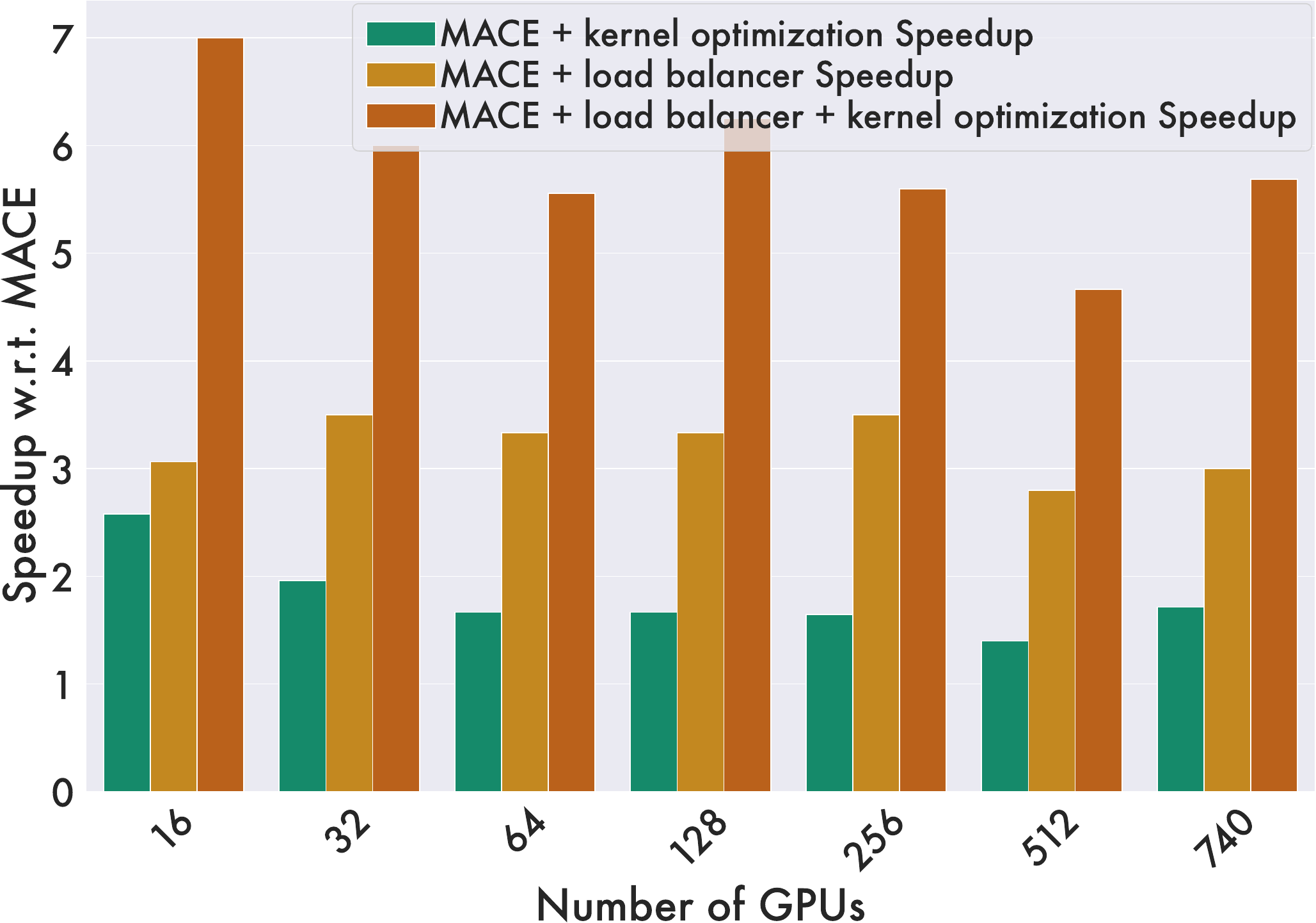}
  \vspace{-0.5em}
  \caption{\small Strong scaling speedup with different optimizations} 
  \label{fig:speedup}
  \vspace{0.5em}
\end{figure}

Our proposed load balancing approach provides a significant improvement over both baseline MACE and kernel optimizations. The kernel optimized MACE consistently outperform their baseline MACE counterparts. The combination of kernel optimization with load balancing provides the best performance across all GPU counts tested, achieving a roughly 6$\times$ speedup over the original MACE at 740 GPUs (185 compute nodes) (\Cref{fig:speedup}). 

Using the formula, \text{strong scaling efficiency} = $\frac{T_1}{P \times T_P} \times 100\%$, where $T_1$ is the execution time on 1 reference number of GPUs, and $T_P$ is the execution time on $P$ GPUs, MACE with load balancer and kernel optimization achieve 
(80/(740/16 × 2)) × 100\% = 86.5\% strong scaling efficiency (calculated from $T_1 \approx$ 80 minutes at 16 GPUs to $T_P \approx$ 2 minutes at 740 GPUs). 

\begin{figure}[t]
  \centering
  \includegraphics[width=0.6\linewidth]{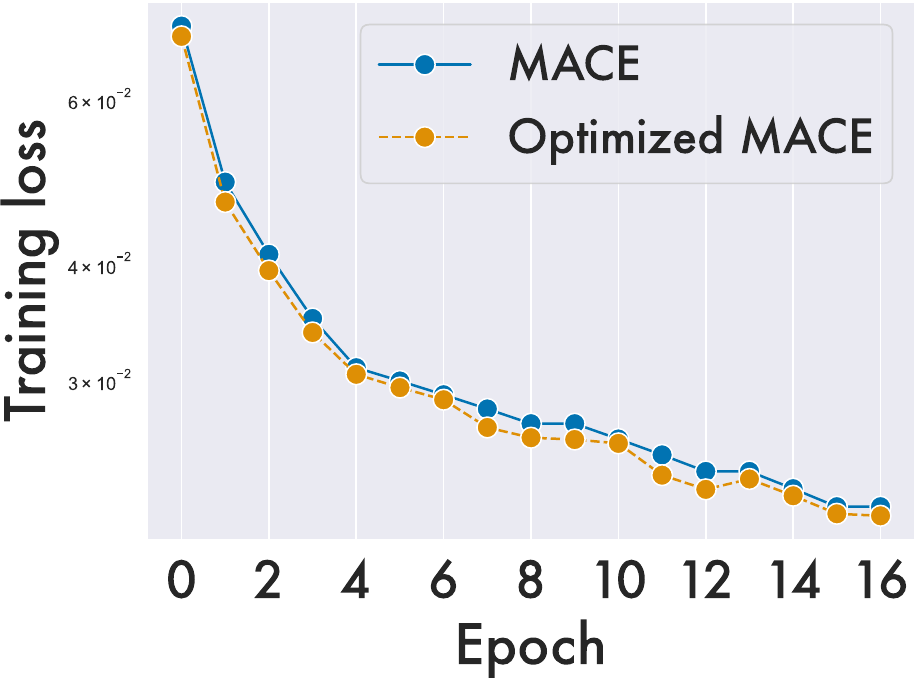}
 \vspace{-1.1em}
  \caption{\small Training loss per epoch for the first 16 epochs.}
  \label{fig:loss_profile}
 \vspace{0.5em}
\end{figure}

The training loss per epoch with the full dataset on 740 GPUs is reported in~\Cref{fig:loss_profile}. The training loss profile of the baseline model and the optimized model follows a similar trajectory (with slight variation in calculated loss), suggesting that the both models evolve their learning in comparable ways.


\begin{figure}[t]
  \centering
  \includegraphics[width=0.8\linewidth]{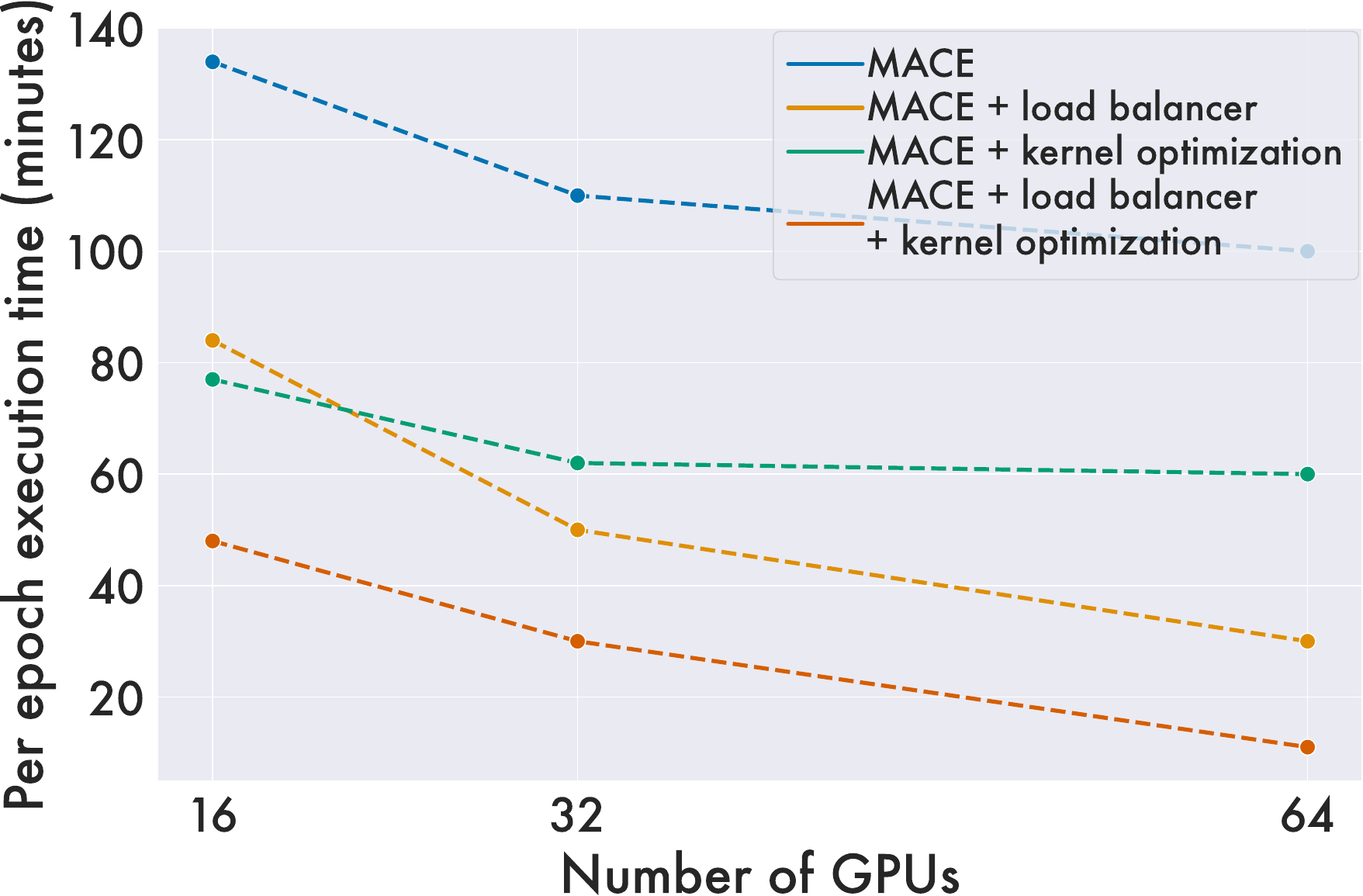}
  \caption{\small Weak scaling results.}
  \label{fig:weak_scaling}
\end{figure}


\subsubsection{Weak scaling}
~\Cref{fig:weak_scaling} shows the weak scaling performance of different MACE configurations, comparing execution time per epoch as the number of GPUs increases while maintaining roughly constant workload (number of graphs) per GPU. The 16-, 32-, and 64-GPU experiments were conducted with the small (0.6 million graphs), medium (1.2 million graphs), and large datasets (2.6 million graphs), respectively.
Load balancing provides significant improvement for both MACE and kernel-optimized MACE. Kernel-optimized MACE with load balancer shows the best weak scaling efficiency, with the smallest increase in time as GPUs increase. 

\subsection{Empirical Determination of Optimal Bin Capacity and Mini-batch Size}\label{sec:bin_capacity_det}
\begin{figure}[t]
  \centering
  \includegraphics[width=0.8\linewidth, height=4.9cm, keepaspectratio]{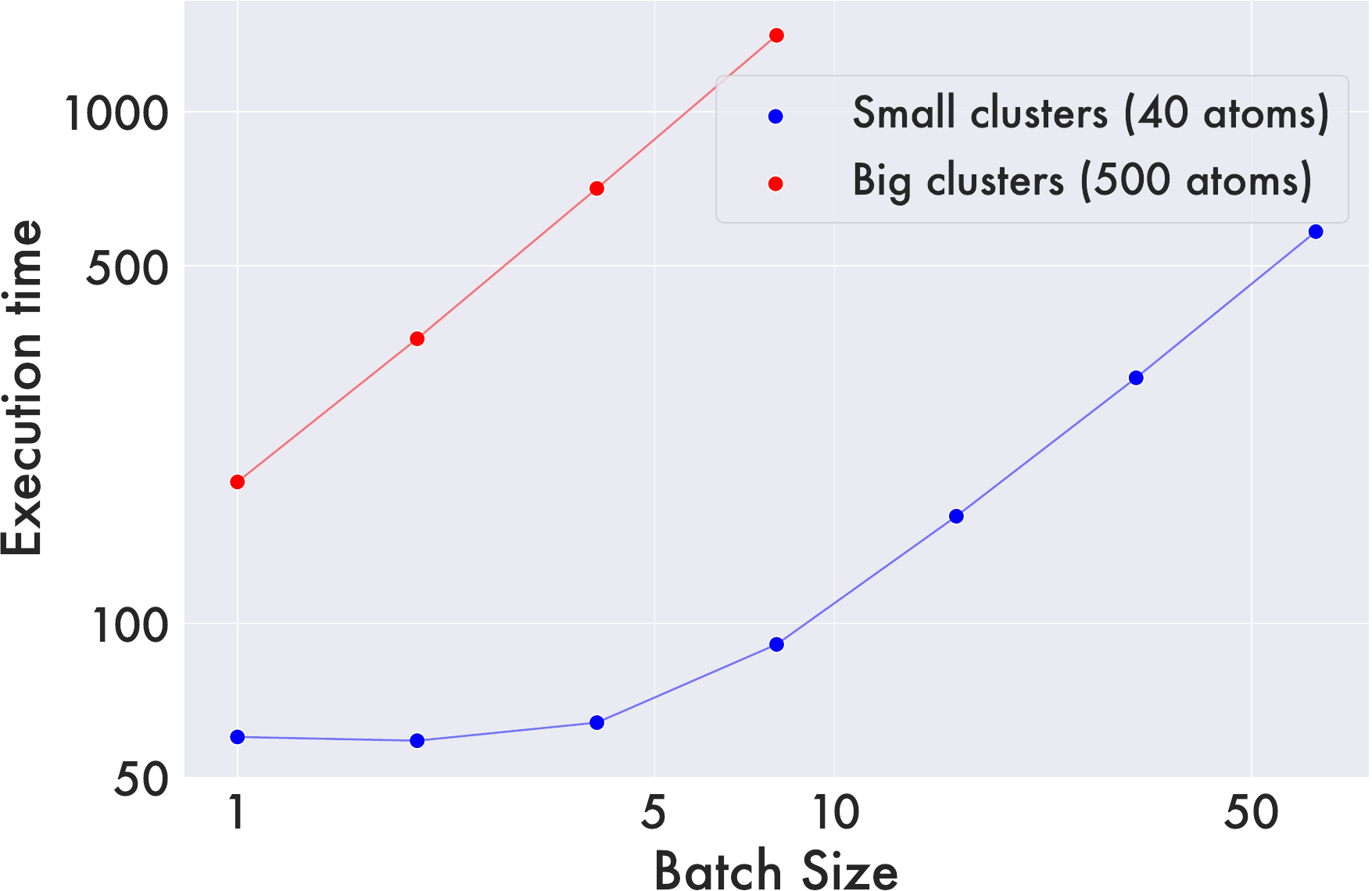} 
  \caption{\small Determining the lower bound on the number of tokens (atoms) in a  batch with \texttt{Float64} data type.}
  \label{fig:lower_bound}
\end{figure}

The capacity of a bin (mini-batch) specified in our load-balancing  algorithm (\Cref{algo:balanced_batches}) is GPU-dependent, with the lower bound determined by the largest graph size in the dataset, and the upper bound limited by GPU memory constraints.
We aim to saturate the compute on each GPU by packing the bins as much as possible, while constrained by the GPU memory capacity. 

Our empirical analysis revealed that, on a single GPU, to saturate the compute, $\approx$ 400 tokens (total atom count i.e. vertex count of 400) and 800 tokens are the lower bound for the number of tokens that will saturate compute when using \texttt{Float64}~(\Cref{fig:lower_bound}) and \texttt{Float32} data types respectively. For example, for \texttt{Float64}, as can be observed from~\Cref{fig:lower_bound}, with large molecules (500 atoms), doubling the batch size also doubles the execution time. With molecules with small number of atoms, the execution time barely changes, suggesting underutilized computational capacity on the GPU, until the batch size reaches 8. The upper bound is capped by the memory ceiling, roughly 2,000 tokens with \texttt{Float64} and 4,000 tokens with \texttt{Float32} settings. Within this range, any value works well, although there is a trade-off to ensure inter-node communication is less frequent while not to have too few batches that may impact the final outcome.

\subsection{Workload Characterization}\label{sec:profiling_workload}
\begin{figure}[t]
 \vspace{-1em}
  \centering
  \includegraphics[width=0.9\linewidth]{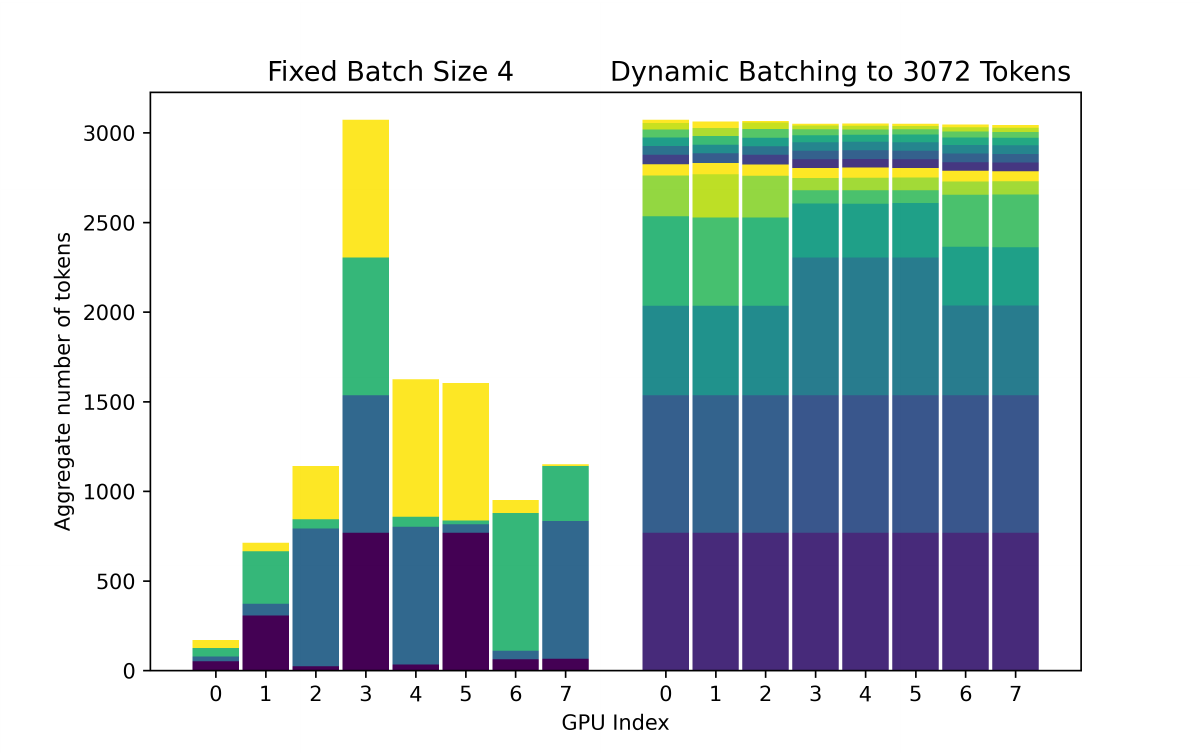}
  \caption{\small Comparison of the distribution of the samples (graphs) with and without the load balancer. Token count refers to the total number of atoms placed on one GPU.}
  \label{fig:batch_imb_lb}
\end{figure}

\begin{figure}
    \centering
    \captionsetup[subfigure]{labelformat=simple, labelsep=colon}
    \renewcommand\thesubfigure{\alph{subfigure}}
    \subfloat[MACE][MACE]{
        \includegraphics[width=0.82\columnwidth, height=15cm, keepaspectratio]{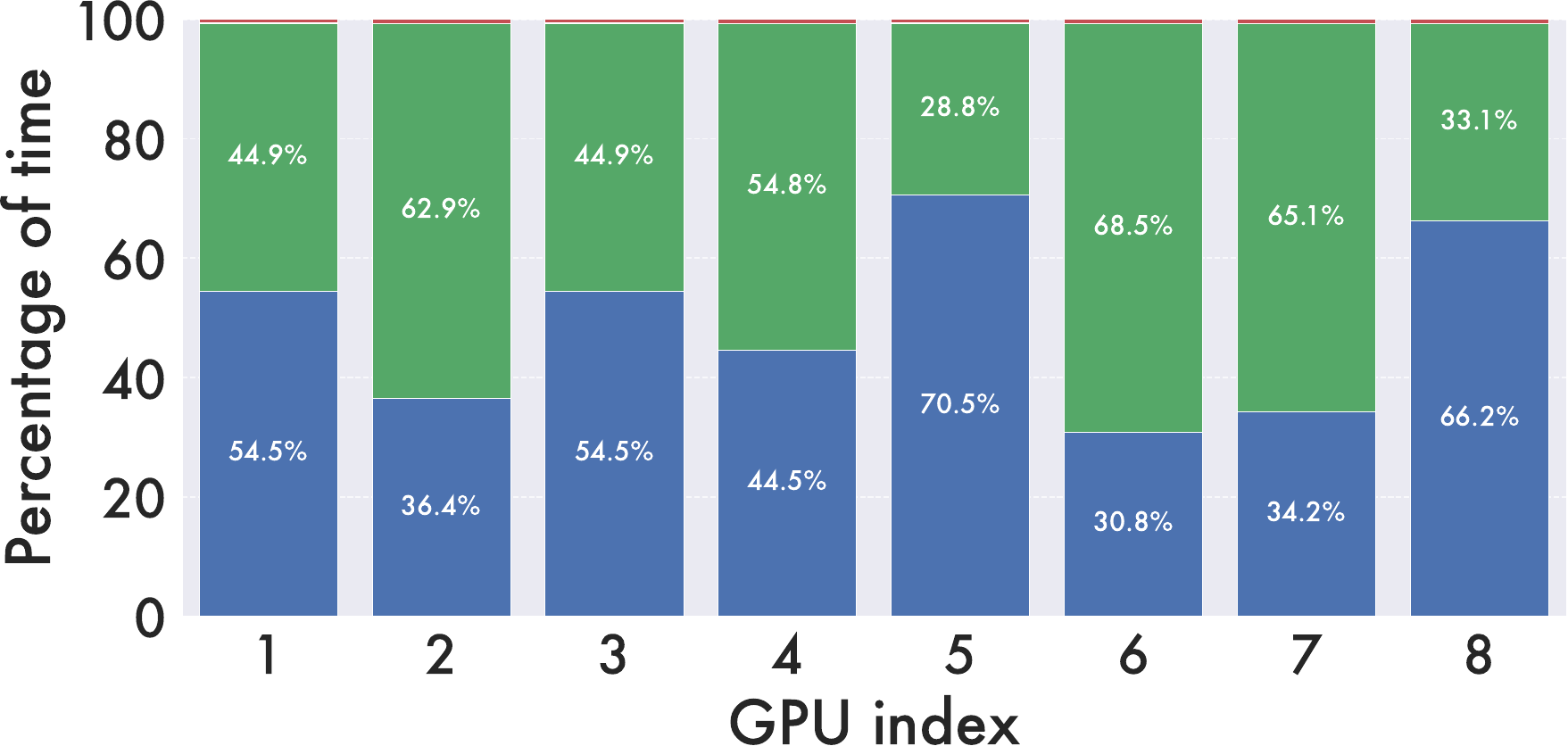}
        \label{fig:breakdown_load_imbalanced}
    }
    \hfill
    \subfloat[Optimized MACE][Optimized MACE]{
        \includegraphics[width=0.82\columnwidth, height=15cm, keepaspectratio]{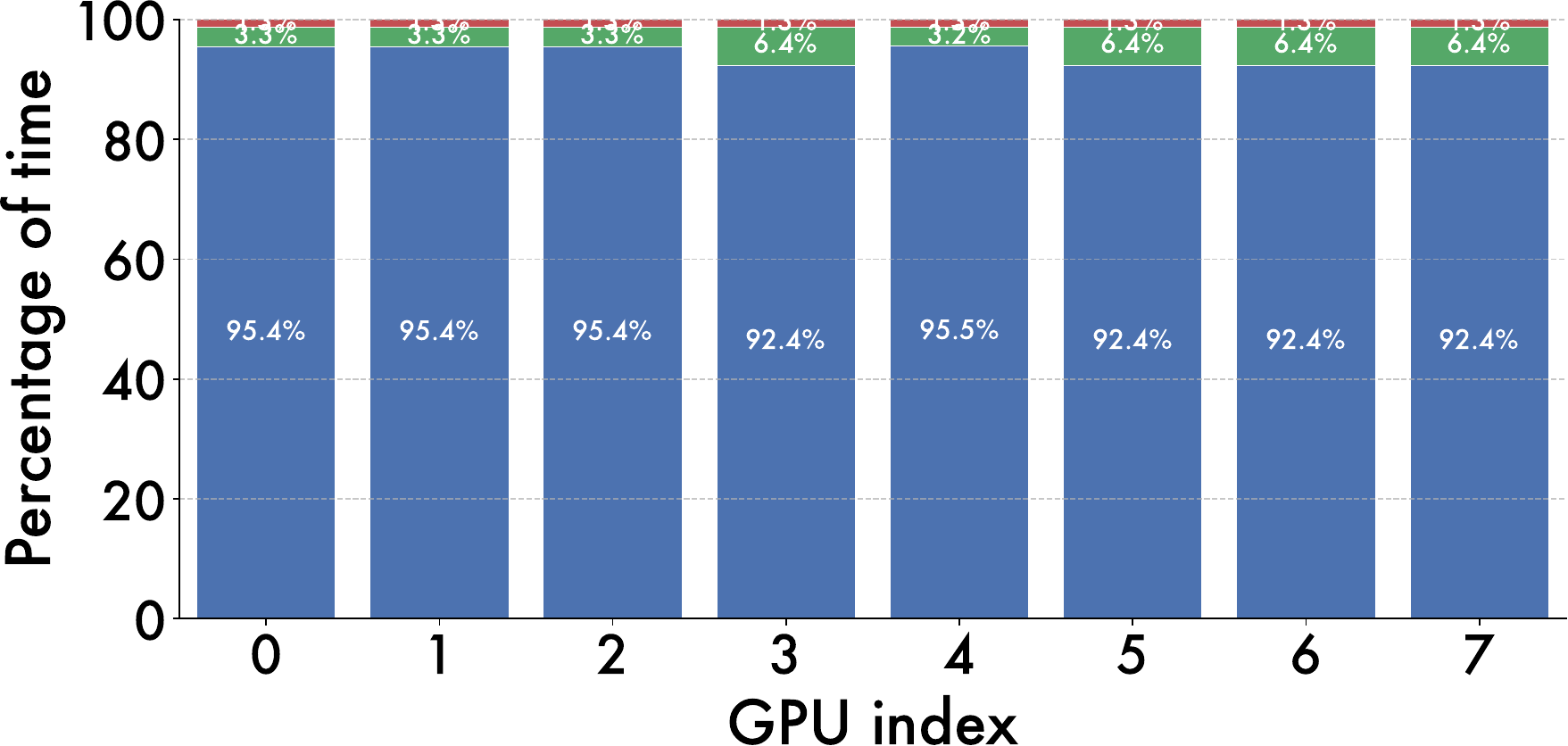}
        \label{fig:breakdown_load_balanced}
    }
    \hfill
    \subfloat{
        \includegraphics[width=0.7\columnwidth]{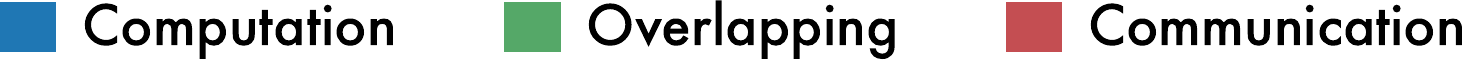}
        \label{fig:legend_breakdown}
    }
    \vspace{-0.9em}
    \caption{\small Computation and communication profiles. Here ``computation'' represents GPU kernel execution time minus overlapping time, while ``overlapping'' indicates the period where computation and communication occur simultaneously.}
    \label{fig:profile}
\end{figure}

Finally, we investigate the impact of data distribution and kernel optimization on  the computation and communication profile of the model. \Cref{fig:batch_imb_lb} provides a snapshot of how graphs are combined in a mini-batch and distributed across 8 GPUs, with the existing approach (left) and with our approach (right). Each color in the figure represents a graph, with the length of each rectangle proportional to the vertex count in that graph. 
Specifically, the left part of the figure shows PyTorch's default \textit{fixed-graph-count distribution} 
 with fixed batch size (graph count) of 4. With this approach, due to the diversity of the vertex count in the molecular graph dataset, each minibatch has widely-varying workload. The runtime is dictated by the straggler on GPU 3. This is evident from \Cref{fig:profile}(a) which demonstrates this imbalance in computation for a fixed batch size.
 
 On the right of \Cref{fig:batch_imb_lb}, we show the distribution of graphs in minibatches with our optimization based approach. Instead of specifying total graph count per mini-batch, we specify a total vertex count (aggregate number of tokens) of 3072 per minibatch (bin), and our algorithm then selects and places the graphs in mini-batches with an objective of balanced data distribution. 
 We observe that our  load balancing approach is able to fit more graphs within the same memory constraint. Notice that with the load balancer, equal distribution of atoms across all GPUs is obtained. 
  More importantly, with load balancing and optimized kernel, the model spends majority of the time in the computation phase (\Cref{fig:profile}(b)), suggesting good throughput, and less time spent in communication. This 
  directly contributes to performance gains.

\section{Related Work}\label{sec:related_work}
This section provides a concise overview of relevant prior work and contextualizes our contribution within the field. 

Allen et al.~present one of the first approaches to developing CFMs for MLIPs in~\cite{allen2023learning}, building on ~\cite{nichol2018first}. The authors aggregate seven datasets calculated with varying quantum mechanical levels of theory to train a CFM for small drug-like molecules ($<$100 atoms). The training process involves repeatedly sampling from disparate datasets and executing a small number of optimization steps 
with each sampled dataset. While this work provided insight for overcoming challenges associated with training CFMs on aggregated datasets, computational challenges associated with large-scale training were not addressed.


Shoghi et al.~introduced an alternative approach, called Joint Multi-domain Pre-Training (JMP) \cite{shoghi2023molecules} that simultaneously trains on multiple datasets from different chemical domains by treating each dataset as a unique pre-training task within a multi-task framework. The authors combined approximately 120 million systems from the OC20, OC22, ANI-1x, and Transition-1x.  While the authors show that fine-tuning the pre-trained model is 12 times faster compared to training from scratch, pre-training was computationally expensive, requiring 34,400 GPU hours for the large model.


Recently, Pasini et al.~introduced the HydraGNN approach to training CFMs over aggregated datasets ~\cite{pasini2024scalable}. HydraGNN uses a multi-headed graph convolutional neural network architecture to employ a Mixture of Experts (MoEs) of 10 models. The authors combined the OC20, OC22, ANI-1x, QM7-X, and MPTrj datasets to obtain approximately 154 million systems and examined a set of invariant and equivariant GNNs.  The study demonstrated near-linear strong scaling performance using more than 2,000 GPUs on the Perlmutter supercomputer and 16,000 GPUs on the Frontier supercomputer.  Load imbalance was identified as a significant challenge, and was attributed to the varying system sizes and was particularly pronounced with equivariant architectures.

\textit{Load imbalance was identified as a significant challenge to scaling performance.} It was attributed to the varying system sizes and was particularly pronounced with equivariant architectures.


\vspace{-0.5em}
\section{Conclusion and Future Work}\label{sec:conclusion} 

This work presents advancements in the training of chemistry foundation models based on graph neural networks, with a particular focus on models employing equivariant operations grounded in spherical harmonics. Molecular datasets exhibit extreme heterogeneity in terms of system sizes, atom types, and sparsity patterns. This diversity, while essential for capturing the complexity of chemical systems, poses unique challenges for efficient parallel training. We make contributions into two crucial areas: data distribution and model training optimization. 

Collectively, these enhancements demonstrate significant performance improvements reducing per-epoch training time from 100 minutes (original MACE) to 18 minutes (optimized MACE) on 64 GPUs, and achieved a 6-fold reduction in per-epoch training time on 740 GPUs from 12 to 2 minutes. Our optimizations target fundamental operations common across equivariant GNN architectures, making them broadly applicable to other models.

One limitation of our developed load balancer is its sacrifice of randomness, which may impact training effectiveness. Additionally, as bottleneck operations in CFMs, such as MACE, are optimized, other operations may become new bottlenecks, requiring further optimization to achieve optimal performance. Regardless of these limitations and necessary future work, the contributions presented here will enable efficient and scalable training of chemistry foundation models.



\bibliographystyle{ACM-Reference-Format}

{
\bibliography{rsc}
}

\end{document}